\documentclass{SciPost}

\hypersetup{
    colorlinks,
    linkcolor={red!50!black},
    citecolor={blue!50!black},
    urlcolor={blue!80!black}
}
\usepackage[utf8]{inputenc}
\usepackage[bitstream-charter]{mathdesign}
\usepackage{subcaption}
\DeclareUnicodeCharacter{2297}{$\scriptstyle\otimes$}
\DeclareUnicodeCharacter{03D5}{$\phi$}

\DeclareSymbolFont{usualmathcal}{OMS}{cmsy}{m}{n}
\DeclareSymbolFontAlphabet{\mathcal}{usualmathcal}

\fancypagestyle{SPstyle}{
\fancyhf{}
\lhead{\colorbox{scipostblue}{\bf \color{white} ~SciPost Physics Codebases }}
\rhead{{\bf \color{scipostdeepblue} ~Submission }}

\fancyfoot[C]{\textbf{\thepage}}
}

\begin{document}

\pagestyle{SPstyle}

\begin{center}{\Large \textbf{\color{scipostdeepblue}{
TensorMixedStates: a Julia library for simulating pure and mixed quantum states using matrix product states
}}}\end{center}

\begin{center}\textbf{
Jérôme Houdayer\textsuperscript{$\star$}
and
Grégoire Misguich\textsuperscript{$\dagger$}
}\end{center}

\begin{center}
Université Paris-Saclay, CNRS, CEA, Institut de Physique Théorique,\\
91191 Gif-sur-Yvette, France
\\[\baselineskip]
$\star$ \href{mailto:jerome.houdayer@ipht.fr}{\small jerome.houdayer@ipht.fr}\\
$\dagger$ \href{mailto:gregoire.misguich@ipht.fr}{\small gregoire.misguich@ipht.fr}
\end{center}

\section*{\color{scipostdeepblue}{Abstract}}
\textbf{\boldmath{%
We introduce TensorMixedStates, a Julia library built on top of ITensor which allows the simulation
of quantum systems in the presence of dissipation using matrix product states (MPS). It offers three key features:
i) it implements the MPS representation for mixed states along with associated operations, in particular the time evolution according to a Lindblad equation or discrete time evolution using non-unitary gates (quantum channels), ii) it is based on ITensor, which has proven its effectiveness and which gives access to efficient low-level tensor manipulation as well as state-of-the-art algorithms (like DMRG or TDVP), finally iii) it presents a
user-friendly interface allowing users to write sophisticated simulations for pure and mixed quantum states in a few lines of code.
}}

\vspace{\baselineskip}



\vspace{10pt}
\noindent\rule{\textwidth}{1pt}
\tableofcontents
\noindent\rule{\textwidth}{1pt}
\vspace{10pt}

\section{Introduction}
\label{sec:intro}

The field  of open quantum many-body problems is a very active area of research in Physics~\cite{breuer_theory_2007}. In the last two decades, there has been huge experimental progress in the
manipulation and in the control of quantum systems such as cold atoms~\cite{blochQuantumSimulationsUltracold2012}, Rydberg atoms~\cite{saffmanQuantumInformationRydberg2010,bernienProbingManybodyDynamics2017,browaeysManybodyPhysicsIndividually2020}, trapped ions~\cite{blattQuantumSimulationsTrapped2012,bruzewiczTrappedIonQuantumComputing2019,monroeProgrammableQuantumSimulations2021}, coupled light-matter systems or superconducting circuits and processors~\cite{devoret_superconducting_2013,ARUTE_QuantumSupremacyUsing_2019} to name a few. Quantum technologies and the development of  devices that are able to perform quantum information tasks~\cite{preskillQuantumComputingNISQ2018} have clearly been a major driving force in this domain. 
These systems are never perfectly isolated from their environment, and the presence of  noise, dissipation and decoherence is often important~\cite{breuer_theory_2007,MORVAN_PhaseTransitionsRandom_2024}. In some situations the presence of the environment can give rise to interesting new phenomena and new dynamical regimes. The environment can even be exploited to engineer useful quantum many-body states~\cite{fazio_many-body_2024}.
Simulating a quantum many-body problem on a classical computer is a notoriously difficult task because the computational cost is in general exponential in the number of constituents and open quantum systems are generally not simpler~\cite{weimer_simulation_2021}. Nevertheless, numerical algorithms where the many-body states are represented (and compressed) using tensor networks 
have established themselves as among the most powerful for this type of problems~\cite{orus_practical_2014,orus_tensor_2019}. Among these methods, those based on matrix-product states (MPS) have proven to be very successful in many situations, and for low-dimensional systems in particular~\cite{schollwock_density-matrix_2011,cirac_matrix_2021}.
In the field of quantum computing, calculations based on MPS have raised the bar concerning the performance that quantum processors must exceed in order to offer a quantum advantage~\cite{zhou_what_2020,ayral_density-matrix_2023,begusic_fast_2024,stoudenmire_opening_2024,king_computational_2024}. In fact, an external environment tends to decrease the amount of entanglement among the degrees of freedom inside the system, and it often
results in a decrease of correlations. This can be a favorable situation for tensor network representations which can exploit the reduced correlations to achieve a better compression of the state.

While there exist several powerful libraries for manipulating pure states with MPS (like ITensor~\cite{itensor} or TenPy~\cite{tenpy2024}),
the software offering for the simulation of open/dissipative systems with MPS is much more limited.\footnote{See for instance \cite{gray2018quimb,pastaq,lacroix_mpsdynamicsjl_2024}.} We attempt here to fill this gap by presenting the TensorMixedStates library~\cite{tensormixedstates} (TMS) which allows one to manipulate {\em mixed} many-body quantum states in the form of MPS. It is based on the ITensor~\cite{itensor} library in Julia and offers a solver for studying the time evolution of open quantum systems described by a Lindblad master equation for the density operator~\cite{Lindblad_generators_1976, gorini_completely_1976,breuer_theory_2007}. It also permits constructing and manipulating density matrices using gates or user-defined quantum channels.
Note that the library does not implement the unraveling of Lindblad master equations into ensembles of quantum trajectories. 

The rest of the paper is organized as follows. We begin in Section~\ref{sec:context} by recalling a few basic notions concerning quantum states and their MPS representation. The end of this section also summarizes the functionalities offered by TMS. 
Section~\ref{sec:features} presents the main features of the TMS library.
This section describes the general design of the library (\ref{ssec:design}), the installation procedure (\ref{ssec:install}), the way to define the Hilbert space (\ref{ssec:space}), the states (\ref{ssec:states}) and the operators (\ref{ssec:operators}). It also presents the main operations that can be performed on these objects: continuous time evolution or discrete gates (\ref{ssec:algo}) and measurements (\ref{ssec:measure}).
Section (\ref{ssec:precision}) presents some functions which help monitor the precision of the simulations, and Section (\ref{sec:high}) illustrates the high-level interface of TMS.

Section~\ref{sec:examples} illustrates the use of TMS with several physical models:
a fermionic chain with dephasing in the bulk (\ref{ssec:fermion_dephasing}), a spin chain with boundary dissipation (\ref{ssec:XX_boundary}), a bosonic chain with incoherent particle injection (\ref{ssec:boson_source}), a fermionic chain with incoherent particle injection (\ref{ssec:fermion_source}),
a model describing the decoherence of a graph state, and a brick wall quantum circuit with noisy gates (\ref{ssec:circuit}). Finally, Section~\ref{sec:conclusion} presents conclusions.

\section{Context}
\label{sec:context}
In this section, we recall a few basic properties and a few notations concerning quantum (pure and mixed) states and their representations with MPS.
\subsection{Pure and mixed quantum states}
The (pure) states of a closed system form a Hilbert space $\cal H$, so that a state $|\psi\rangle\in\cal H$ is a vector.
Given an operator acting on $\cal H$ there are essentially three basic operations that we may consider: i) measuring the expectation value
of an observable $O$ (with $O$ hermitian)
\begin{equation}
  \langle O\rangle = \langle\psi| O |\psi\rangle,
\end{equation}
ii) doing some discrete evolution by applying a gate $U$ (with $U$ unitary)
\begin{equation}
  |\psi\rangle = U |\psi_0\rangle,
\end{equation}
or iii) doing continuous-time evolution with the Hamiltonian $H$ ($H$ hermitian)
\begin{equation}
  \partial_t |\psi\rangle = -i H |\psi\rangle.
\end{equation}

For an open system, this formulation is no longer sufficient, and a state must be represented as
a density matrix $\rho$ which must be Hermitian, positive semidefinite with unit trace~\cite{breuer_theory_2007}.
When the state is pure, $\rho$ is simply a projector
\begin{equation}
  \rho = |\psi\rangle\langle\psi|
\end{equation}
but for general mixed states we have ${\rm Tr}\left[\rho^2\right]<1$.
The three operations mentioned above for pure states become
\begin{align}
  \langle O\rangle &= \mathrm{Tr}(O \rho),\\
  \rho &= U \rho_0 U^\dagger,\\
  \partial_t \rho &= -i [H, \rho],
\end{align}
where $[A, B]=AB-BA$ is the commutator. For a mixed state, a general discrete evolution is a  linear and completely positive map (also called quantum operation, or quantum channel) and takes the form~\cite{nielsen_quantum_2010}
\begin{equation}
  \rho = \sum_i E_i \rho_0 E_i^\dagger,
\end{equation}
with Kraus operators $\left\{E_i\right\}$ satisfying $\sum_i E_i^\dagger E_i = \mathbb{1}$. In a continuous-time context, the evolution, if Markovian, can instead be modelled by the Gorini–Kossakowski–Sudarshan–Lindblad master equation~\cite{gorini_completely_1976,Lindblad_generators_1976}
\begin{equation}
  \partial_t \rho = {\cal L}(\rho),
\end{equation}
where $\cal L$ is the Lindbladian
\begin{equation}
  {\cal L}(\rho) = -i [H, \rho] + \sum_k\left(L_k \rho L_k^{\dagger}-\frac{1}{2}\{L_k^{\dagger}L_k, \rho\}\right),
\end{equation}
with no particular constraints on the $L_k$ and $\{ A , B \}=AB+BA$ is the anti-commutator.

\subsection{Matrix product states and operators}
In quantum many-body problems the dimension $d$ of $\cal H$ grows exponentially with the system size and this severely limits
the sizes accessible to numerically exact calculations. A possible option is then to use approximate representations of the states, and MPS~\cite{schollwock_density-matrix_2011} provide such approximate representations.
Suppose $\cal H$ is a tensor product of $N$ finite-dimensional local Hilbert spaces ${\cal H}_i$ of dimension $d_i$: ${\cal H} = {\cal H}_1 \otimes {\cal H}_2 \otimes \cdots \otimes {\cal H}_N$ (thus $d = d_1 d_2\cdots d_N$). The system consists of $N$ sites, each associated with a local Hilbert space of dimension $d_i$. In the pure case, any state $|\psi\rangle$ can be written
\begin{equation}
  |\psi\rangle = \sum_{\{s\}}T_1^{s_1} T_2^{s_2} \cdots T_N^{s_N}|s_1 s_2 \cdots s_N\rangle,
  \label{eq:psi-mps}
\end{equation}
where $T_i^{s_i}$ is
a $\chi_{i-1}\times \chi_i$ matrix and $s_i$ runs over all single site basis states of ${\cal H}_i$ ($\chi_0$ and $\chi_N$ are set to 1). $T_i$ as a whole can be seen as a 3-index tensor of dimension $\chi_{i-1}\times \chi_i \times d_i$ or alternatively as a $\chi_{i-1}\times \chi_i$ matrix whose elements are quantum states belonging to ${\cal H}_i$, in which case we can simply write
\begin{equation}
  |\psi\rangle = T_1 T_2 \cdots T_N,
\end{equation}
hence the name matrix product state.

For large systems, the {\em exact} representation of a generic state as an MPS requires most of the bond dimensions $\chi_i$
to grow exponentially with $N$. MPS are particularly useful when there exist matrices $T_i$ of size
much smaller than $d$ which provide a good approximation of the target state $|\psi\rangle$.
In practice, we set a maximum bond dimension $\chi$, and we numerically approximate the states of interest by an MPS with $\chi_i\leq \chi$.
The larger $\chi$ is, the larger the precision of the approximation is.

To operate on an MPS, we can build a matrix-product operator (MPO) 
\begin{equation}
  O = O_1 O_2 \cdots O_N,
\end{equation}
where the $O_i$ are matrices whose elements are operators acting on ${\cal H}_i$, that is the $O_i$
are tensors with four indices (two of which having dimension $d_i$) and we then have
\begin{equation}
  O |\phi\rangle = (O_1\cdot T_1)(O_2\cdot T_2)\cdots(O_N\cdot T_N).
\end{equation}

\begin{figure}
\centering
\subcaptionbox{}{\includegraphics[width=0.3\textwidth]{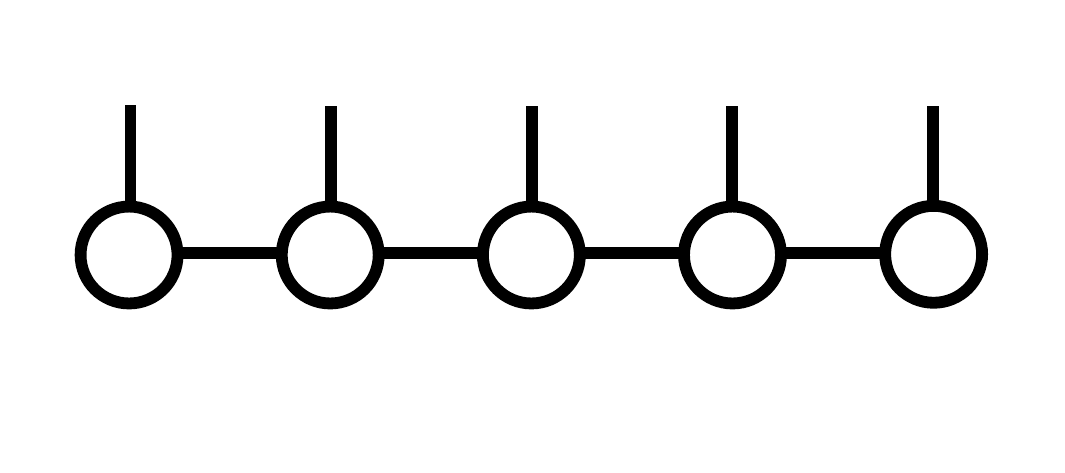}}
\subcaptionbox{}{\includegraphics[width=0.3\textwidth]{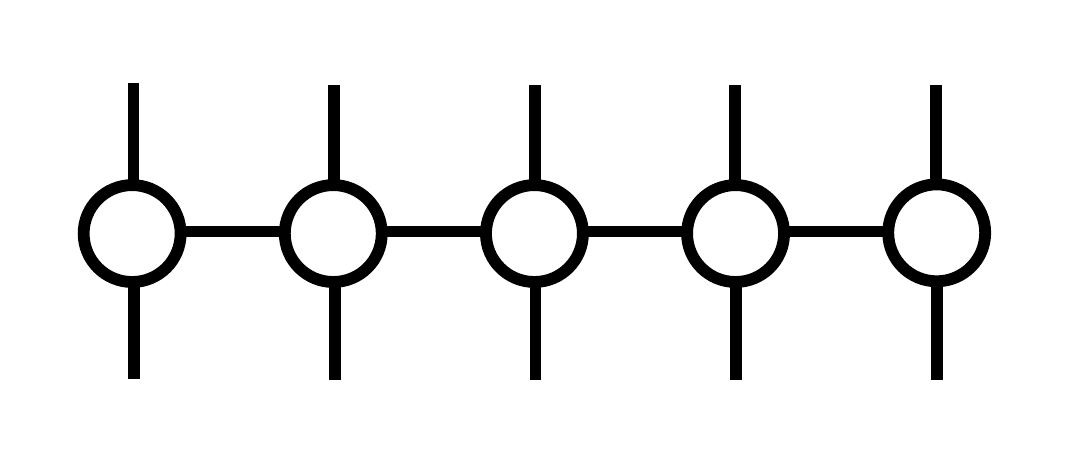}}

\subcaptionbox{}{\includegraphics[width=0.3\textwidth]{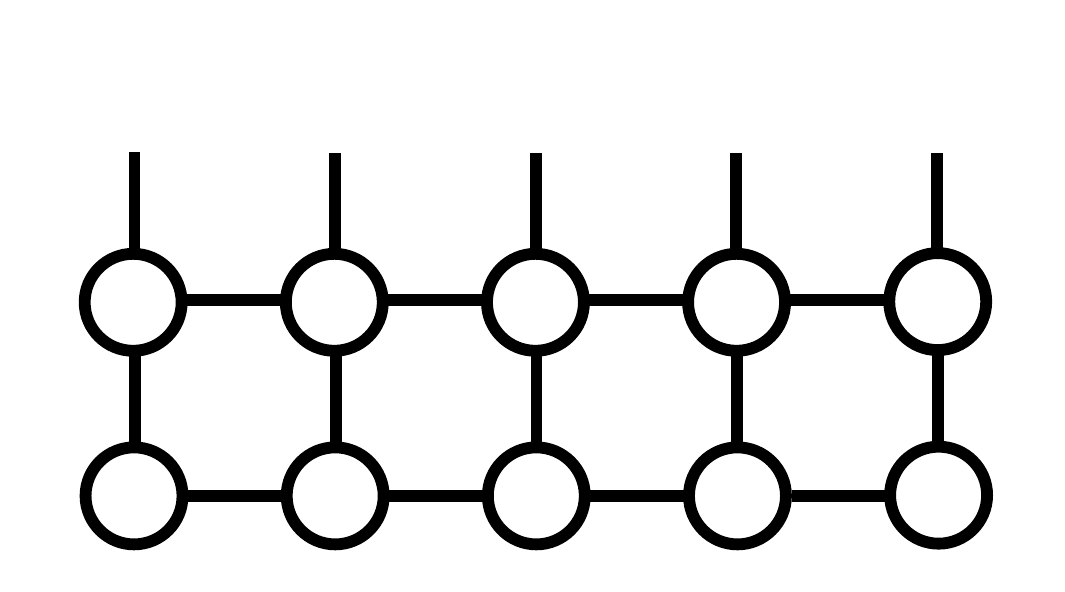}}
\subcaptionbox{}{\includegraphics[width=0.3\textwidth]{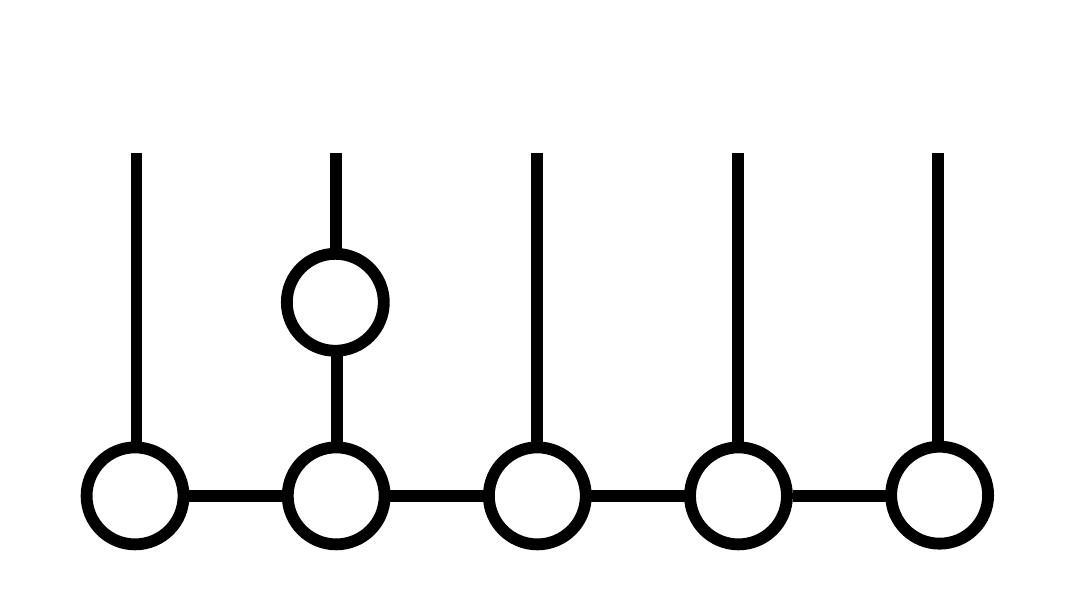}}
\subcaptionbox{}{\includegraphics[width=0.3\textwidth]{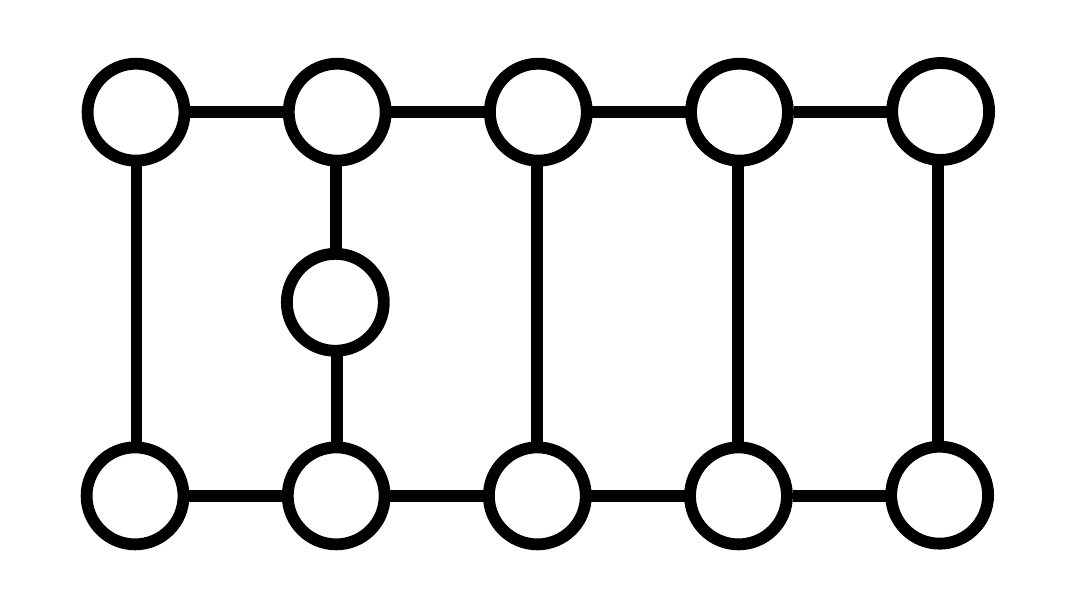}}
\caption{Tensor diagrams for pure states: tensors are represented by circles and indices by lines. A contraction is represented by a line (index) connecting two circles (tensors).
(a) MPS with five sites. (b) MPO with five sites. (c) application of an MPO to an MPS. (d) Application of a single site operator. (e) Mean value of a single site operator (MPS represented with downward indices are complex conjugated).}
\label{fig:diag_pure}
\end{figure}

Simple operators, like single site operators, do not need to be represented as MPOs. They can be directly applied to the corresponding $T_i$, this is very useful to compute mean values of operators.  All this can be nicely represented by tensor diagrams as shown on Fig.~\ref{fig:diag_pure}

What about mixed states? We can see the density matrix $\rho$ of a mixed state as a vectorized state $|\rho\rangle\rangle$ in a larger space of dimension $d^2$~\cite{breuer_theory_2007} and apply the same process as before and write
\begin{equation}
  \rho = \sum_{\{s\}, \{s'\}}R_1^{s_1, s'_1} R_2^{s_2, s'_2} \cdots R_N^{s_N, s'_N}
  |s_1 s_2 \cdots s_N\rangle\langle s'_1 s'_2 \cdots s_N|,
  \label{eq:rho-vect}
\end{equation}
where $R_i^{s_1, s'_1}$ is a $\chi_{i-1}\times \chi_i$ matrix and $s_i$ and $s'_i$ run over all single site basis states of ${\cal H}_i$. $R_i$ as a whole is then a 4-index tensor of dimensions
$\chi_{i-1}\times \chi_i \times d_i \times d_i$. In the vectorization process
${\cal H}_i \otimes{\cal H}_i$ is seen as the space of dimension $d_i^2$ of the operators acting on ${\cal H}_i$, and we can also write
\begin{equation}
  |\rho\rangle\rangle = \sum_{\{o\}}R_1^{o_1} R_2^{o_2} \cdots R_N^{o_N}
  |o_1 o_2 \cdots o_N\rangle\rangle,
\end{equation}
where $o_i$ runs over all single site operator basis of ${\cal H}_i$
\footnote{In the simplest case where the system is made of qubits ($d_i=2$), the $o_i$ can be chosen as the identity and the three Pauli matrices.} and $R_i$ can be seen as a 3-index tensor of dimension $\chi_{i-1}\times \chi_i \times d_i^2$. We can finally see $R_i$ as $\chi_{i-1}\times \chi_i$ matrix whose elements belong to ${\cal H}_i \otimes{\cal H}_i$ and simply write
\begin{equation}
  |\rho\rangle\rangle = R_1 R_2 \cdots R_N.
\end{equation}

\begin{figure}
\centering
\subcaptionbox{}{\includegraphics[width=0.3\textwidth]{diag_pure_mpo.pdf}}
\subcaptionbox{}{\includegraphics[width=0.3\textwidth]{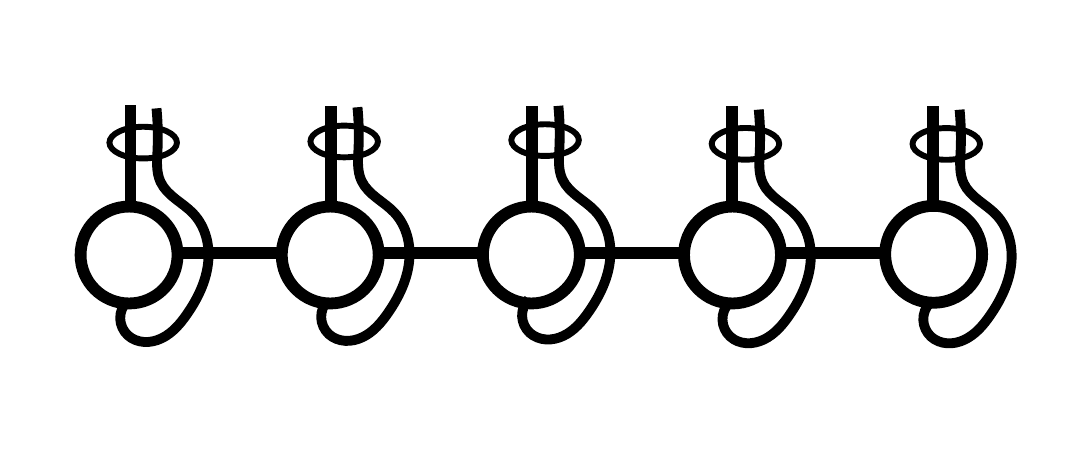}}
\subcaptionbox{}{\includegraphics[width=0.3\textwidth]{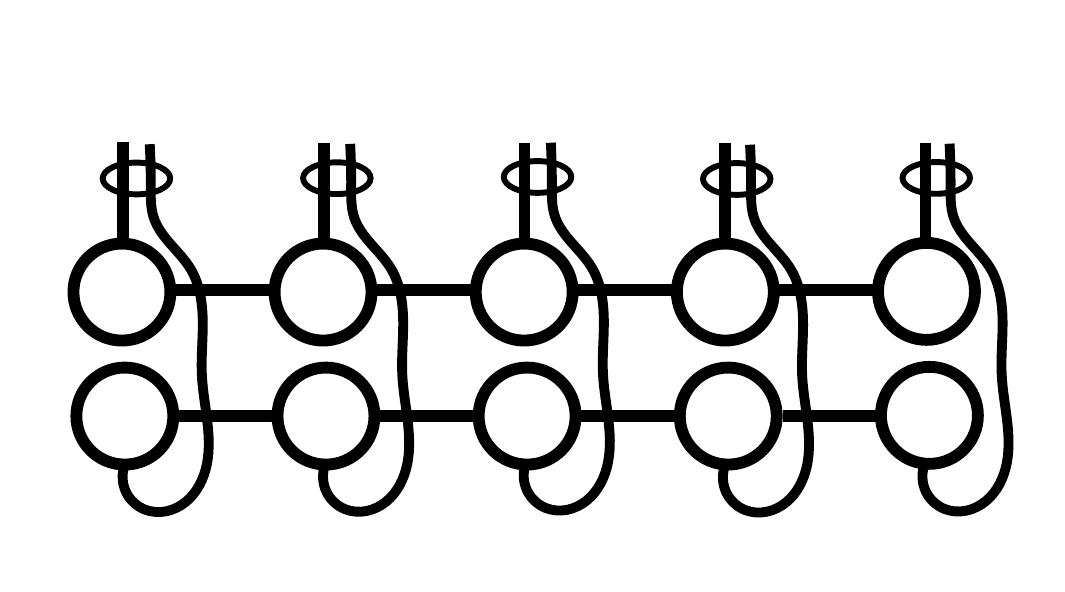}}

\subcaptionbox{}{\includegraphics[width=0.3\textwidth]{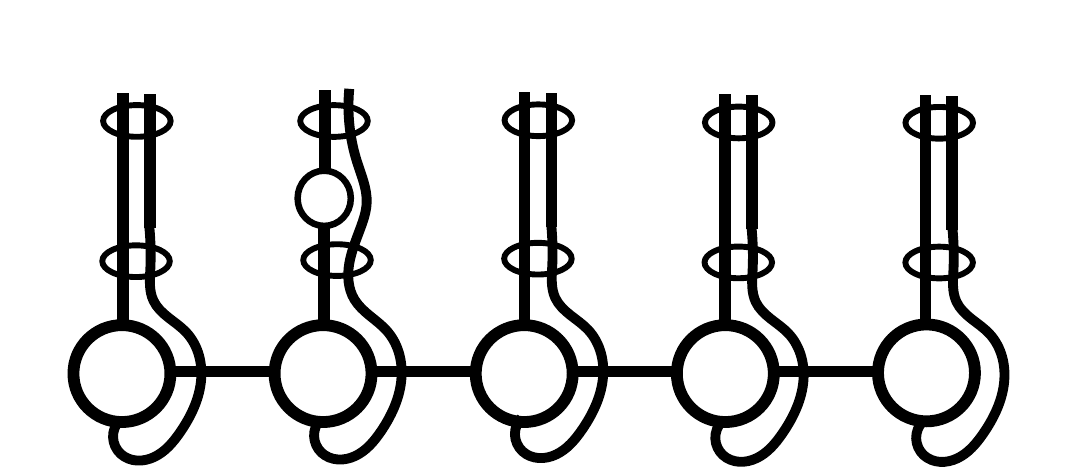}}
\subcaptionbox{}{\includegraphics[width=0.3\textwidth]{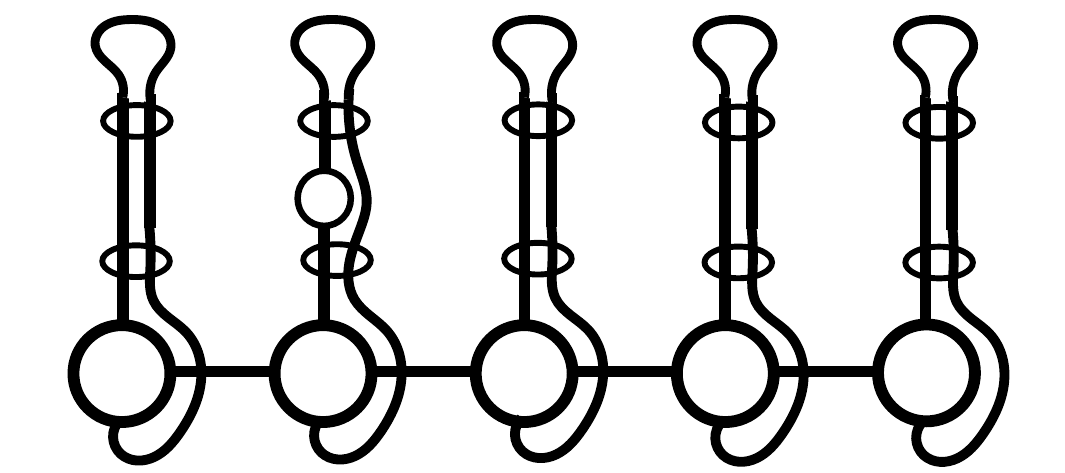}}
\caption{Tensor diagrams for mixed states: two lines surrounded by a circle represent combined indices, that is two indices combined as one.
(a) Density matrix $\rho$ with five sites represented as an MPO. (b) Vectorized density matrix $|\rho\rangle\rangle$ represented as an MPS. (c) Vectorized density matrix $|\rho\rangle\rangle$ of a pure state. (d) Application of a single site operator $O$. (e) Mean value ${\rm Tr}\left[\rho O\right]$ of a single site operator.}
\label{fig:diag_mixed}
\end{figure}

Note that, contrary to the matrix-product density operator representation~\cite{verstraete_matrix_2004}, the representation
of Eq.~\ref{eq:rho-vect} does not guarantee that $\rho$ is positive. In practice, this is not an issue as long as the bond dimension can be taken large enough to ensure a sufficient accuracy for the observables of interest.

For a pure state $|\psi\rangle$, we have $\rho = |\psi\rangle\langle\psi|$ and thus
$R_i^{s_i, s'_i} = T_i^{s_i}\otimes T_i^{s'_i}\dagger$ or correspondingly $R_i = T_i\otimes T_i^\dagger$. Note that in this operation $\chi_i$ for $R$ is the square of $\chi_i$ for $T$.

If one needs to represent an operator acting on a density matrix $\rho$, one may use the above strategy with an MPO acting on the vectorized state $|\rho\rangle\rangle$.
The $O_i$ are then matrices whose elements are operators acting on the ``doubled'' Hilbert space. The $O_i$ can also be viewed as tensors with four indices, two of which having dimension $(d_i)^2$. Tensor diagrams for mixed states are shown on Fig.~\ref{fig:diag_mixed}.

An important question is the size of the matrices  that are required to represent accurately a given state as an MPS. For a pure state $|\psi\rangle$, it is well known that the dimension $\chi$ of the bond which separates the left region $A$ from the right region $B$\footnote{In an MPS a given matrix index naturally separates the system in two subsystems.} has to be scaled as the exponential of the bipartite von Neumann entropy $S_{\rm vN}^{A-B,|\psi\rangle}$ associated to the bipartition. In other words $\ln \chi \sim S_{\rm vN}^{A-B}$.
For a mixed state $\rho$ and a given bipartition $A-B$ of the system, one can consider the (vectorized) pure state $|\rho\rangle\rangle$ and the von Neumann entropy $S_{\rm vN}^{A-B,|\rho\rangle\rangle}$ associated to the bipartition. The operator space entanglement entropy (OSEE) associated to the $A-B$ partition
is, by definition, equal to this entanglement entropy~\cite{prosen_operator_2007}: $S_{\rm OSEE}^{A-B} = S_{\rm vN}^{A-B,|\rho\rangle\rangle}$.
When approximating $\rho$ by an MPS as in Eq.~\ref{eq:rho-vect}, the required bond dimension $\chi$ must then scale as the exponential of the OSEE, $\ln \chi \sim S_{\rm OSEE}^{A-B}$.

\subsection{Functionalities offered by TensorMixedStates}

The ITensor library provides tools for constructing and manipulating MPSs and MPOs. In particular, it provides the following four main operations: (i) creation of an MPS state representing a pure product state, (ii) creation of an MPO representing an operator acting on pure states, (iii) measurement of the mean value of an operator on a pure state, (iv) powerful algorithms for evolving a state with an operator: in particular applying the operator on the state (MPO / MPS contraction), computing the ground state of an operator (DMRG~\cite{white_density_1992,schollwock_density-matrix_2011}), and solving a continuous equation of the form $\partial_t |\psi\rangle = O |\psi\rangle$ (TDVP~\cite{yang_time-dependent_2020}).

Unfortunately, ITensor has not been developed with mixed states in mind and while the evolution algorithms (point (iv)) are quite general and can be used in the mixed case, the other three components are not (points (i), (ii) and (iii)). Moreover, even if ITensor is largely extendable, using this extendability to accommodate the new needs was at best awkward, in particular it would have been complicated to use the operator framework of ITensor.

TMS thus relies on the powerful core algorithms of ITensor (tensor contractions, DMRG and TDVP) and reimplements from scratch all the rest to accommodate mixed state representations. Namely, it provides (i) the creation of an MPS representing a mixed product state, (ii) the creation of an MPO representing an operator acting on a mixed state, (iii) measurement of the mean value of an operator on a mixed state.

Recall that in the mixed-state setting there is a wider variety of operator actions to consider. In the pure-state case, an operator 
$O$ is naturally applied to a state $|\psi\rangle$ by computing $O |\psi\rangle$. By contrast, when working with a density matrix 
$\rho$, an operator $O$ can act in several distinct ways, namely $O\rho$, $O\rho O^\dagger$, $[O, \rho]$ and $\{O^\dagger O, \rho\}$. It is this richness that led us to develop a complete and flexible operator framework.

Finally, TMS also provides algorithms not present in ITensor, namely the W$^{\rm I}$ and W$^{\rm II}$ approximations~\cite{zaletel_time-evolving_2015} that build an MPO for approximately representing $\exp(t O)$ at small $t$.
These time-evolution algorithms are complementary to TDVP, in particular for non-Hermitian evolutions where TDVP is sometimes less efficient (see for instance Sec.~\ref{ssec:fermion_source}). TMS also provides an algorithm to compute the steady state of a Lindbladian $\cal L$, simply computing the ground state of $\cal L^\dagger \cal L$ with DMRG.

\section{Features}
\label{sec:features}
\subsection{Design choices}
\label{ssec:design}
TMS is the successor of the Lindbladmpo library~\cite{landa_nonlocal_2023,lindbladmpo}.
Lindbladmpo was based on the C++ version of the ITensor library and since the ITensor library has migrated to the Julia language, Lindbladmpo could not benefit
from the recent developments of ITensor. In addition, Lindbladmpo was limited to qubits (two-dimensional local Hilbert space on each site) while TMS is much more general.

In this context, we decided to create a more general and more flexible software to address the simulation of
open quantum systems with non-unitary evolution. The choice of the Julia language was then natural to ease the interactions
with ITensor. Moreover, using Julia helped us develop a more flexible and more user-friendly interface.
Finally, the combined use of Julia and ITensor brings multithreading for free.

The main decision concerned the use of the operator and state framework of ITensor. As stressed in the previous section, the difficulty lies
in the fact that, for a given operator, one has four different possible actions on $\rho$. 
It proved challenging to use ITensor's native structures for that, and we decided to write from scratch a new state and operator framework. This also required writing new functions for creating MPS and MPO from states and operators.
While this amounted to a substantial and bookkeeping-heavy implementation effort, it did not involve
any choice of nontrivial algorithm or any decision concerning approximation schemes.
All algorithmic decisions are delegated to ITensor itself: TMS primarily acts as a wrapper that formats data appropriately for ITensor and forwards the user-specified parameters.
Likewise for W$^{\rm I}$ and W$^{\rm II}$ approximations, we implemented the algorithms as described in~\cite{zaletel_time-evolving_2015,bidzhiev_out--equilibrium_2017}. Finally, the steady-state computation routine is simply a three-line wrapper around an ITensor DMRG call.

We finally decided on a double interface: (i) a high-level interface allows one to design state-of-the-art simulations
in a few lines of code as demonstrated in Sec.~\ref{sec:high}. (ii) a low-level interface gives access to all
the features of the library as we now explain.

\subsection{Installation and usage}
\label{ssec:install}
To use the TMS library, one must first add it to the Julia environment. This is done the usual way in Julia by
\begin{verbatim}
  ]add TensorMixedStates
\end{verbatim}
Then, to use it in a script, one has to add a "using" clause at the top of the file
\begin{verbatim}
  using TensorMixedStates
\end{verbatim}

Note that the examples shown in this article, together with other examples, are available with the source on GitHub~\cite{tensormixedstates}. Moreover,
full documentation is also available online (see the README.md file~\cite{tensormixedstates}).

\subsection{Hilbert space}
\label{ssec:space}

The first step in a quantum simulation consists in describing the Hilbert space.
In our case, the Hilbert space must be a finite product over $N$ local finite-dimensional Hilbert subspaces.
Those local subspaces can be chosen independently among the "site types" proposed by TMS. At the moment, there
are seven possible choices: qubit, boson, fermion, spin, electron, tJ and q-boson. This set is easily extendable
(see TMS online documentation for details).

In TMS, the Hilbert space is described with a \verb"System" object
\begin{verbatim}
  # a system with 10 qubits
  sys1 = System(10, Qubit())

  # a hybrid system with 3 sites
  sys2 = System([Qubit(), Boson(4), Fermion()])
\end{verbatim}
Note that for bosons it is necessary to truncate the Fock space and the dimension $d$ of the local Hilbert space
must be provided as a parameter. In the example above, the second site has $d=4$, which corresponds to
a maximal occupation boson number equal to $d-1=3$.

\subsection{States and representations}
\label{ssec:states}

The starting point for building states are single site states. For each kind of site, predefined states designated by
their names are defined. For example, for qubits one has "Up" or "0", "Dn" or "1" and "+" and "-" and more (for a list of all predefined state names, see the online documentation). A state which is not predefined can be represented by a vector of complex numbers such as \verb"[1, 0]". For mixed representations, there is one predefined state for all site types called "FullyMixed" which corresponds to the infinite temperature state (with $\rho = \mathbb{1} / d$ proportional to the identity matrix). A general local mixed state can be represented by a matrix of complex numbers such as $[[0.5\ 0];[0\ 0.5]]$.

In TMS, both pure and mixed representations are held by \verb"State" objects. All operations on
\verb"State" objects will work in the same way, with the same syntax independently of the nature, pure or mixed,
of the state (as long as such operations make sense for this representation).

We can build a product state (i.e. $|\psi\rangle = |\psi_1\rangle \cdots|\psi_N\rangle$) with the \verb"State" constructor
parametrized by \verb"Pure" or \verb"Mixed" and two arguments: the system and a vector of single site states (either defined by their names or explicitly given as vectors or matrices), for the usual case where all single site states are the same, we can simply give that state. For example
\begin{verbatim}
  # an all up pure state for a 10 qubit system
  st1 = State{Pure}(sys1, "Up")

  # a mixed state for a hybrid system
  st2 = State{Mixed}(sys2, [[1, im]/sqrt(2), "1", "FullyMixed"])
\end{verbatim}
As a shortcut, one can define both system and state in one call, for example
\begin{verbatim}
  # an up and down state for a 5 qubit system
  st1 = State{Pure}(5, Qubit(), ["Up", "Dn", "Up", "Dn", "Up"])

  # an infinite temperature state for a hybrid system
  st2 = State{Mixed}([Qubit(), Boson(4), Fermion()], "FullyMixed")

\end{verbatim}
In this case the system can be retrieved from the state by accessing the \verb"system" field of the state.

To build more complicated states, one can form linear combinations of states. For example, one can build the
GHZ state on the 10 qubit system with
\begin{verbatim}
  ghz = (State{Pure}(sys1,"Up") + State{Pure}(sys1,"Dn")) / sqrt(2)
\end{verbatim}

Many functions are defined for simple tasks on \verb"State" objects,
for example to get some information on the state: \verb"length" to get the number of sites,
\verb"maxlinkdim" to get the maximum bond dimension, \verb"trace" and \verb"trace2" to get
$\mathrm{Tr}(\rho)$ and $\mathrm{Tr}(\rho^2)$ and so on.
To obtain the mixed state representation of a pure state one can use
\begin{verbatim}
  mixed_state = mix(pure_state)
\end{verbatim}

\subsection{Operators}
\label{ssec:operators}

In TMS, operators are a key feature. They allow the user to represent any kind of quantum operator to operate on the state.
They come in two flavors, {\em generic} (like \verb"X") or {\em indexed} (like \verb"X(3)"). The main difference is that indexed
operators are applied on specific sites of the system whereas generic operators are not. For example, \verb"Z" is the Pauli operator
$\sigma^z$ and \verb"Z(3)" is the Pauli operator $\sigma^z$ applied on site number 3.

For each site type several commonly used operators are predefined.
For example, for the qubit space, \verb"X, Y, and Z" are the Pauli operators, \verb"Sp" is the raising operator $S^+$, \verb"H" is the Hadamard gate, \verb"Swap" is the Swap gate, \verb"Id" is the identity operator.
For the boson space, \verb"A" is the annihilation operator, \verb"dag(A)" is the creation operator and \verb"N" is the number operator.
See the online documentation for an exhaustive list of predefined operators.

Suppose that \verb"A" and \verb"B" are operators
and that \verb"x" is a real or complex number and \verb"i" and \verb"j" integers.
Then the following operations are defined:
\verb"x*A", \verb"A/x": multiplication by a scalar,
\verb"A+B", \verb"A-B": sum of operators, \verb"A*B": product of operators, \verb"A^x": power,
\verb"exp(A)": exponentiation, \verb"dag(A)": dagger, \verb"A⊗B", \verb"tensor(A, B)": tensorial product (the symbol $\otimes$ is usually obtained by typing \verb"\otimes" in a Julia aware editor), \verb"A(i)", \verb"A(i, j)": indexation, \verb"controlled(A)": controlled gates for qubits (see below), \verb"Dissipator(A)": Lindblad dissipator operator (see below) and \verb"Gate(A)": the gate operator (see below).

Hamiltonians can thus be written in one line, for example
\begin{equation}
  H = -J \sum_{i=1}^{n-1} X_i X_{i+1} + Y_i Y_{i+1},
\end{equation}
is built by
\begin{verbatim}
  hamiltonian = -j * sum(X(i)X(i+1)+Y(i)Y(i+1) for i in 1:n-1).
\end{verbatim}
Here we have used the Julia \verb"sum" function that adds the given iterator using \verb"+" and the compact Julia product syntax
(we can omit "*" in some cases).

One can define Lindblad dissipators with the \verb"Dissipator" function. For example, \verb"Dissipator(Sp)" represents 
the jump operator toward up for a qubit and
\begin{verbatim}
  dissipators = gamma * (Dissipator(Sp)(1) + Dissipator(Sp)(N))
\end{verbatim}  
is the sum of two jump operators. It can describe incoherent spin flips on the boundary sites of a qubit chain, at a rate $\gamma$.
By convention the Lindbladian is written 
\begin{verbatim}
  lindbladian = -im * hamiltonian + dissipators.
\end{verbatim}
The \verb"hamiltonian" term in the above Lindbladian definition represents the (super)operator $\rho\to [H,\rho]$
and each term $L_k$ in \verb"dissipators" represents the (super)operator
$\rho\to L_k \rho L_k^\dagger - \frac{1}{2}\{L_k^\dagger L_k, \rho\}$. In the example above, the two dissipators correspond to
$\rho\to \gamma \left(S_1^+ \rho S_1^- - \frac{1}{2}\{S_1^- S_1^+, \rho\}\right)$ and
$\rho\to \gamma \left(S_n^+ \rho S_n^- - \frac{1}{2}\{S_n^- S_n^+, \rho\}\right)$.

One can define noisy gates (quantum channels) with the \verb"Gate" function, for example
\begin{verbatim}
  K = 0.9 * Gate(Id) + 0.1 * Gate(X)
\end{verbatim}
defines an operator which acts as follows on a mixed state $\rho$:
\begin{equation}
  K \rho = 0.9 \rho + 0.1 X\rho X.
\end{equation}
One can define more complex operators such as
\begin{equation}
  R_{xy}(\phi) = \exp \left[-i{\frac {\phi }{4}}(X\otimes X+Y\otimes Y)\right],
\end{equation}
by simply typing the following 
\begin{verbatim}
  Rxy(t) = exp(-im * t / 4 * (X ⊗ X + Y ⊗ Y)).
\end{verbatim}
Another example is the Toffoli gate, which can be constructed by
\begin{verbatim}
  toffoli = controlled(controlled(X))
\end{verbatim}
Finally, one can also define new operators by giving their matrix with
\begin{verbatim}
  Iswap = Operator{2}("Iswap", [1 0  0 0
                                0 0 im 0
                                0 im 0 0
                                0 0  0 1]).
\end{verbatim}
In the instruction above \verb"{2}" is the number of sites on which the operator acts.

\subsection{Algorithms}
\label{ssec:algo}

There are four main algorithms that one can use in TMS. First,
one can apply an operator $O$ as a quantum gate: $|\psi\rangle \to O |\psi\rangle$ for pure states and
$\rho \to O \rho O^\dagger$ for mixed states. This is implemented by
\begin{verbatim}
  new_state = apply(my_gate, old_state; options...),
\end{verbatim}

One can also perform time-evolutions. This can be done with a Schrödinger evolution $\partial_t |\psi\rangle = -i H |\psi\rangle$ for pure state, or a with Lindblad evolution for mixed states. The library provides two algorithms to do this, one using TDVP (called \verb"tdvp")~\cite{yang_time-dependent_2020} and one
using the W$^{\rm I}$ or W$^{\rm II}$ (called \verb"approx_W") MPO approximation (see Zaletel {\it et al.}~\cite{zaletel_time-evolving_2015}). The syntax is as follows:
\begin{verbatim}
  lindbladian = -im * hamiltonian + dissipators
  new_state = tdvp(-im * hamiltonian, t, old_state; options...)
  new_state = tdvp(lindbladian, t, old_state; options...)
  new_state = approx_W(-im * hamiltonian, t, old_state; options...)
  new_state = approx_W(lindbladian, t, old_state; options...)
\end{verbatim}
where \verb"t" is the integration time and the \verb"options" allow one to set parameters such as the integration time step, the level of truncation,
or the definition of intermediate measurements (via objects called {\it observers} similar to those of ITensor).

Note that \verb"tdvp" and \verb"approx_W" can be used with time-dependent Hamiltonians and/or dissipators and that the \verb"approx_W" schemes are available up to the order 4 in the time step $\tau$ (leading to a Trotter error which scales as $\mathcal{O}(\tau^5)$~\cite{bidzhiev_out--equilibrium_2017}).

Another algorithm is the computation of the ground state using DMRG \cite{white_density_1992,schollwock_density-matrix_2011}. It is essentially the algorithm from the ITensor library:
\begin{verbatim}
 energy, ground_state = dmrg(hamiltonian, start_state; options...)
\end{verbatim}

Finally, it is also possible to use DMRG for mixed states in order to minimize the "square" $\cal L^\dagger \cal L$ of the Lindbladian.
This allows one to compute directly the steady-state of the system without the need to perform a long time evolution.\footnote{The steady state is the eigenstate of $\cal L$ associated to the eigenvalue 0, hence it is the zero-energy ground state of the Hermitian operator $\cal L^\dagger \cal L$.} This is achieved by
\begin{verbatim}
  energy, s_state = steady_state(lindbladian, start_state; options...)
\end{verbatim}

A more thorough explanation of the available options for the different algorithms
can be found in the online documentation~\cite{tensormixedstates}.

\subsection{Measurements}
\label{ssec:measure}

Finally, we describe how to obtain the expectation value of an observable on a \verb"State". For example, one writes
\begin{verbatim}
  measure(state, X(1))
\end{verbatim}
A more complicated example would be
\begin{verbatim}
  measure(state, 0.5X(1)Z(3)-im*X(2)Y(4))
\end{verbatim}
One can also  ask for the set of expectation values $\langle X(1)\rangle\cdots\langle X(N)\rangle$ of an operator $X$ on all sites
\begin{verbatim}
  measure(state, X)
\end{verbatim}
or even a correlation matrix
\begin{verbatim}
  measure(state, (X, Y))
\end{verbatim}
One can also use predefined functions on \verb"State" like \verb"Trace" for the trace or \verb"Linkdim"
for the bond dimension (this set of functions is extendable by the user). One can also make several measurements in a single call
\begin{verbatim}
  measure(state,  [X(1), X, (X, Y), Purity, Linkdim])
\end{verbatim}

\subsection{Following computation precision}
\label{ssec:precision}

As with any simulation software, it is important to assess the precision of a computation. Here in TMS, the representation of mixed states does not guarantee that the density matrix stays Hermitian positive with trace one. While this allows more computational deviations to happen (due to truncations or Trotter errors for instance), it is possible to monitor such deviations. Even if we do not have access to the eigenvalues of the density matrix to check the positivity, we can monitor its trace and its hermiticity. These quantities can in turn be used to adjust the parameters of the simulation (time step, truncation threshold and maximal bond dimension).

There are four predefined observables to do this:  \verb"Trace", \verb"TraceError", \verb"Hermiticity" and \verb"HermiticityError". The hermiticity is between 0 and 1, 0 for anti-Hermitian and 1 for Hermitian. The error versions measure the deviation from the correct value (which is 1).

TMS also provides the \verb"Trace2" observable which computes $\mathrm{Tr}(\rho^2)$, the purity of the state.
Note that if the accumulated errors become really important the presence of spurious negative eigenvalues in $\rho$ can lead to
a purity that is apparently larger than 1.

Note that TMS works perfectly with non-normalized density matrices; observables are then correctly normalized using the actual trace.

\subsection{High-level interface}
\label{sec:high}

The high-level interface of the TMS library is accessed via the function \verb"runTMS".
It can be used alone or together with the low-level interface for finer control. The goal of the high-level interface is to be able to design fully fledged simulations with minimal code. All the examples presented in the next section were created using this interface.

The principle is the following: define a sequence of actions to be applied on a state and pass it to \verb"runTMS".
As a first example, consider the following complete script
\begin{verbatim}
using TensorMixedStates, .Fermions

hamiltonian(n) = -sum(dag(C)(i)C(i+1)+dag(C)(i+1)C(i) for i in 1:n-1)
dissipators(n, gamma) = sum(Dissipator(sqrt(4gamma) * N)(i) for i in 1:n)

sim_data(n, gamma, step) = SimData(
    name = "Fermion tight-binding chain with dephasing noise",
    phases = [
        CreateState(
            type = Mixed(),
            system = System(n, Fermion()),
            state = [ iseven(i) ? "Occ" : "Emp" for i in 1:n ]),
        Evolve(
            duration = 4,
            time_step = step,
            algo = Tdvp(),
            evolver = -im*hamiltonian(n) + dissipators(n, gamma),
            limits = Limits(cutoff = 1e-30, maxdim = 100),
            measures = [
                "density.dat" => N,
                "OSEE.dat" => EE(div(n, 2))
            ]
        )
    ]
)

runTMS(sim_data(40, 1., 0.05))
\end{verbatim}

The first line brings our library in scope and the definitions for the fermion site type.
The next two lines define the evolution operator (more details on the model in Sec.~\ref{ssec:fermion_dephasing}).
Then we have a function definition for \verb"sim_data" to build a \verb"SimData" object
corresponding to the parameters of the simulation. The \verb"name" field in \verb"SimData"
sets the name of the directory where the output will be stored. The \verb"phases" field describes
the simulation. Here there are two phases: first create a state, second compute the time evolution. The field names
are self-explanatory. The \verb"measures" field describes the data files created by the simulation (here there are two files \verb"density.dat"
and \verb"OSEE.dat") and what quantities are to be written in each of them (here the mean fermion occupancy measured on each site and the operator-space entanglement entropy (OSEE)
at the middle point of the chain). Note that one can put more than one observable per file.
The last line calls \verb"runTMS" with the chosen parameters.

In this case, \verb"runTMS" will create a directory with the given name, run the simulation and put the results
in the corresponding files. In addition to the data files, it will also create a "log" file to follow the progress
of the simulation and register information and warnings, it will also copy the program script to \verb"prog.jl"
for information and reproducibility.

In addition to \verb"CreateState" and \verb"Evolve" there are other available phases:
\verb"ToMixed" to turn a pure representation into a mixed representation, \verb"Dmrg"
to use the DMRG algorithm, \verb"Gates" to apply gates, \verb"PartialTrace" to trace out some sites and finally \verb"SteadyState" to compute the steady state of a Lindbladian.

As already stated, more detailed information on the syntax and options can be found in the online documentation~\cite{tensormixedstates}.

\section{Examples}
\label{sec:examples}
This section illustrates the use of the TMS library to study several dissipative
quantum problems evolving according to a Lindblad equation.
The examples are chosen because they have been the focus of recent studies in the literature and because an exact solution
is available.\footnote{Except for the model in Sec.~\ref{ssec:circuit}.} These solutions allow checking quantitatively the numerical results. The first example is a fermionic
chain with dephasing noise (Sec.~\ref{ssec:fermion_dephasing}), the second example is a spin chain with boundary
dissipation (Sec.~\ref{ssec:XX_boundary}),  the third example is a one-dimensional bosonic model with an incoherent
particle source in the center of the chain (Sec.~\ref{ssec:boson_source})
and the fourth example (Sec.~\ref{ssec:fermion_source}) is the fermionic version of the previous bosonic model.
The fifth example (Sec.~\ref{ssec:complete_graph}) describes the dissipation of a complete-graph state in a qubit system.
Finally, the model of Sec.~\ref{ssec:circuit} is a deep quantum circuit with unitary 2-qubit gates as well as dissipative channels which model qubit errors.

A runnable and well commented version of the code of these examples is available in the \verb"examples/article" section of the repository on GitHub~\cite{tensormixedstates}.

\subsection{Fermion tight-binding chain with dephasing noise}
\label{ssec:fermion_dephasing}

\begin{figure}
  \centering
  \includegraphics[width=0.9\textwidth]{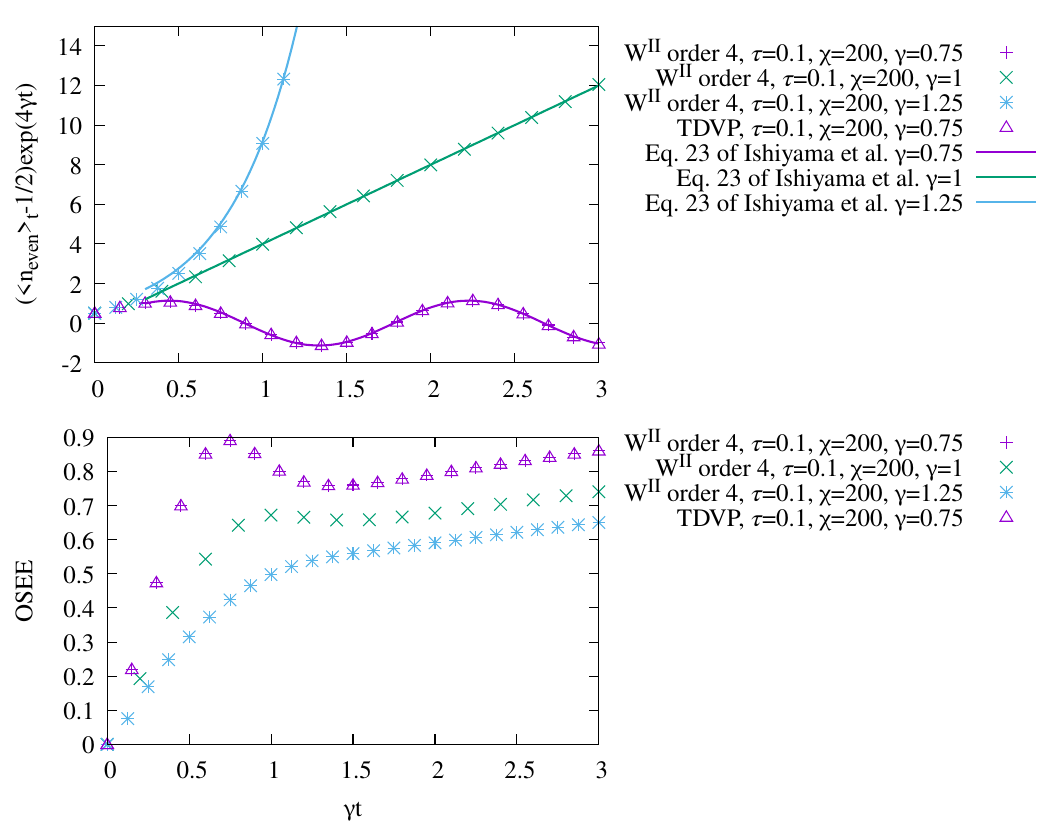}
  \caption{Top: fermion density $\langle n_i\rangle$(t) on even sites $i$ as a function of time in a tight binding chain with dephasing noise and three different strengths of the noise ($\gamma=0.75,1.0,1.25$). The density is evaluated by averaging over 4 sites in the center of the system. Solid lines indicate the exact asymptotic results of Ref.~\cite{ishiyama_exact_2025}.  Bottom: OSEE as a function of time, computed for the bipartition in the middle of the chain. The simulations have been carried out with the W$^{\rm II}$ algorithm at order 4 and with TDVP (see legend).}
  \label{fig:fermion}
\end{figure}

We consider here the one-dimensional spinless fermion model studied in \cite{ishiyama_exact_2025}. The initial state is a pure state where the even sites are occupied, and the odd ones are empty. The system then evolves under the action of a nearest-neighbor hopping Hamiltonian
\begin{equation}
  H=-\sum_{i=1}^{N-1} \left(c^\dagger_i c_{i+1}+c^\dagger_{i+1} c_i \right)
\end{equation}
as well as under the following "dephasing" jump operators:
\begin{equation}
  L_i=\sqrt{4\gamma} n_i,
\end{equation}
where $n_i=c^\dagger_i c_i$ is the fermion number operator on site $i$.
Using the high-level interface, we combine the Hamiltonian and the jump operators into an 
\verb"evolver" (see Sec.~\ref{sec:high} for the full listing)
\begin{verbatim}
hamiltonian(n) = -sum(dag(C)(i)C(i+1)+dag(C)(i+1)C(i) for i in 1:n-1)
dissipators(n, gamma) = sum(Dissipator(sqrt(4gamma)*N)(i) for i in 1:n)
\end{verbatim}
Dissipative problems where the Hamiltonian and the Lindbladian are quadratic in the creation and annihilation operators can be solved exactly~\cite{znidaric_exact_2010,eisler_crossover_2011,zunkovic_closed_2014,barthel_solving_2022}. The present model was recently studied in the limit of an infinite chain~\cite{ishiyama_exact_2025}. The authors of this study demonstrated that the system displays oscillatory decay or over-damped decay, depending on the strength $\gamma$ of the dissipation. In Fig.~\ref{fig:fermion}, numerical results for the fermion density, obtained with the present library, are compared to the exact asymptotic results derived from~\cite{ishiyama_exact_2025}.
The fermion density converges to $1/2$ at long times. To magnify how such convergence occurs the vertical axis represents deviation from $1/2$ multiplied by an exponential factor $\exp(4\gamma t)$. We observe a good agreement between the simulations and the analytical behavior derived in \cite{ishiyama_exact_2025}.

The bottom panel of Fig.~\ref{fig:fermion} represents the evolution of the OSEE~\cite{prosen_operator_2007,znidaric_complexity_2008} associated with the central bipartition of the chain. The OSEE quantifies the total amount of correlations between the two subsystems. This quantity is very useful in the context of MPS since the bond dimension $\chi$ (on the bond associated to the bipartition) required to represent $\rho$ faithfully is expected to
obey a scaling of the form $\ln(\chi) \sim {\rm OSEE}$.

\subsection{{XX} spin chain with boundary dissipation}
\label{ssec:XX_boundary}

We illustrate here the use of the TMS library to simulate the dynamics of an open spin-$\frac{1}{2}$ chain with Lindblad terms acting at its boundaries.
The Hamiltonian is the so-called XX model
\begin{equation}
  H=\sum_{i=1}^{N-1} \left(\sigma^x_i \sigma^x_{i+1} + \sigma^y_i \sigma^y_{i+1} \right),
  \label{eq:H_XX}
\end{equation}
and the dissipation is due to four Lindblad operators acting at both ends of the chain:
\begin{eqnarray}
	&&L_1=\sqrt{\varepsilon_{\mathrm{L}}\frac{1+\mu_{\mathrm{L}}}{2}}\sigma_1^+,\quad L_3=\sqrt{\varepsilon_{\mathrm{R}}\frac{1+\mu_{\mathrm{R}}}{2}}\sigma_N^+,\label{sourcedissipation} \\
	&&L_2=\sqrt{\varepsilon_{\mathrm{L}}\frac{1-\mu_{\mathrm{L}}}{2}}\sigma_1^-,\quad L_4=\sqrt{\varepsilon_{\mathrm{R}}\frac{1-\mu_{\mathrm{R}}}{2}}\sigma_N^-,\label{sinkdissipation}.
  \label{eq:L_XX}
\end{eqnarray}
 $\sigma^{\pm}=(\sigma^x\pm i\sigma^y)/2$, $\varepsilon_{\rm{L,R}}$ are the strengths of the coupling between the spin chain and the reservoirs at both ends. $\mu_{\rm{L,R}}$ are the magnetization of each reservoir. Coding such model with TMS can be done by defining the following \verb"evolver":
 \begin{verbatim}
  hamiltonian(n) = sum(X(i)*X(i+1)+Y(i)*Y(i+1) for i in 1:n-1)
  dissipators(n, eL, muL, eR, muR) =
    Dissipator(sqrt(eL*(1+muL)/2)*Sp)(1) +
    Dissipator(sqrt(eL*(1-muL)/2)*Sm)(1) +
    Dissipator(sqrt(eR*(1+muR)/2)*Sp)(n) +
    Dissipator(sqrt(eR*(1-muR)/2)*Sm)(n)
 \end{verbatim}

\begin{figure}
  \centering
  \includegraphics[width=0.8\textwidth]{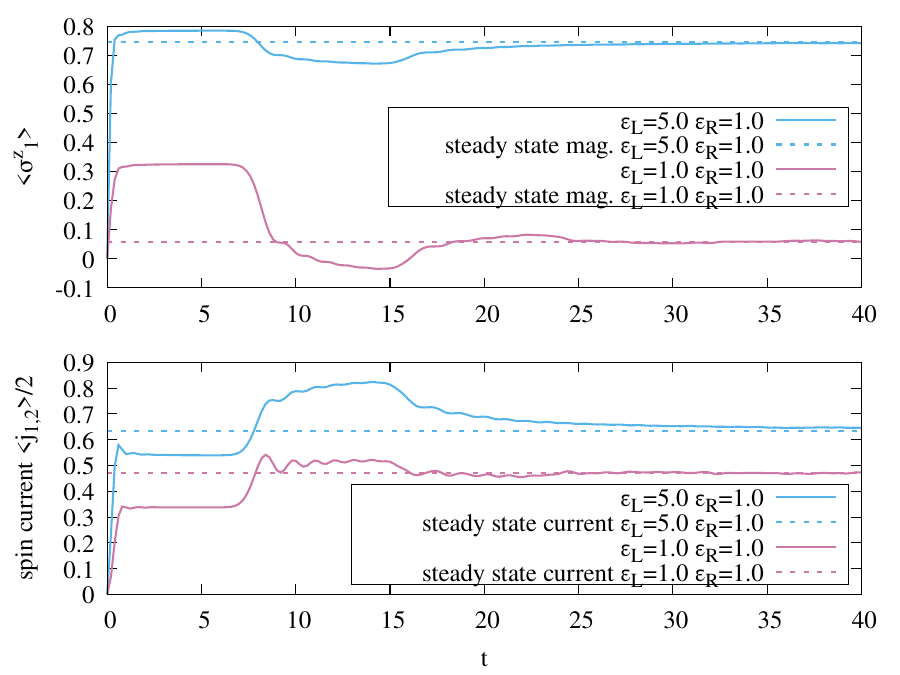}
  \caption{XX spin chain with boundary dissipation (Eqs.~\ref{eq:H_XX}-\ref{eq:L_XX}) Top: mean magnetization $\langle \sigma^x_1\rangle$ on the first spin as a function of time. Bottom: mean spin current $\langle \sigma^x_1 \sigma^y_2- \sigma^x_1 \sigma^y_2\rangle$ on the first bond. Physical parameters: infinite-temperature initial state, system size $N=30$, $\mu_{\mathrm{L}}=-\mu_{\mathrm{R}}=1.0$,  $\varepsilon_{\mathrm{R}}=1.0$.
  Green curves: $\varepsilon_{\mathrm{L}}=5.0$, red curves:  $\varepsilon_{\mathrm{L}}=1.0$.  
  Simulation parameters: maximum bond dimension $\chi=300$, time step $\tau=0.1$, algorithm: W$^{\rm II}$ at order 4.  
  These curves should be compared with Fig.~4 of \cite{yamanaka_exact_2023}.}
  \label{fig:XX}
\end{figure}

Via the Jordan-Wigner transformation the Hamiltonian above maps to a quadratic fermionic Hamiltonian and the Lindblad terms become linear in the fermionic creation and annihilation operators. This model is thus said to be quasi-free\cite{prosen_third_2008,barthel_solving_2022} and can be solved exactly. The dynamics of this model was studied in Ref.~\cite{yamanaka_exact_2023} (see also \cite{znidaric_exact_2010} for the exact steady-state). Fig.~\ref{fig:XX} displays the time evolution of two observables: the mean magnetization $\langle \sigma^x_1\rangle$ on the first spin and the mean spin current $\langle \sigma^x_1 \sigma^y_2- \sigma^x_1 \sigma^y_2\rangle$ on the first bond. The data is in agreement with the results of Ref.~\cite{yamanaka_exact_2023}.

\subsection{Free bosons with a localized source}
\label{ssec:boson_source}

\begin{figure}
  \centering
  \includegraphics[width=0.8\textwidth]{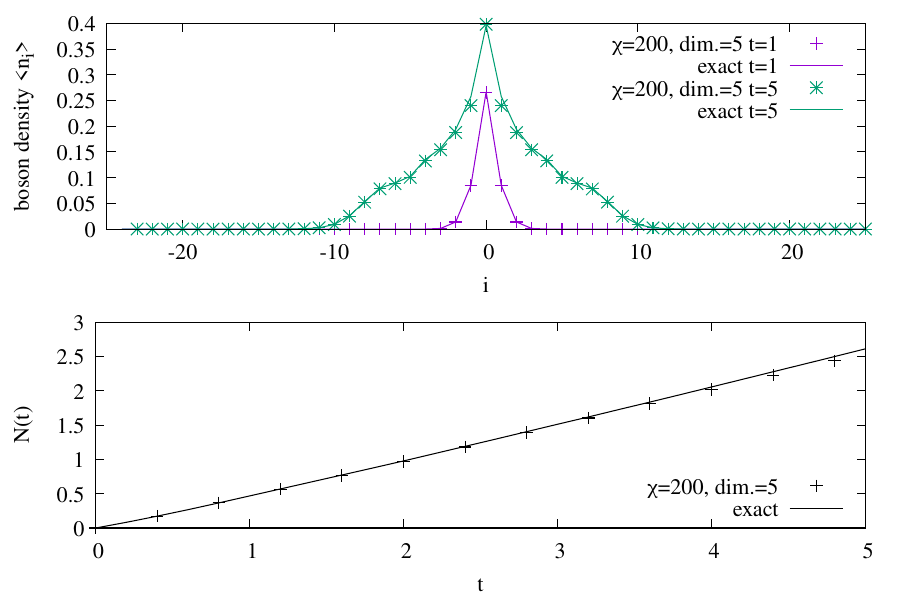}
  \caption{Free boson model with a localized source (Eqs.~\ref{eq:HBS}-\ref{eq:LBS}).
  Top: density profile $\langle b^\dag_i b_i \rangle$
  Bottom: mean total number of bosons $N(t)$, numerics versus exact result.
  Physical parameters: system size $N=50$, $\Gamma=0.2$.
  Simulation parameters: maximum bond dimension $\chi=200$, local dimension: dim=5 (boson occupancy $\leq4$), time step $\tau=0.1$, algorithm: W$^{\rm II}$ at order 4. 
  }
  \label{fig:boson}
\end{figure}

We show here how the library can be used to simulate a dissipative system with bosonic degrees of freedom.
The unitary part of the dynamics is generated by a free (quadratic) boson Hamiltonian on a chain:
\begin{equation}
H =  \sum_{i=-N/2+1}^{N/2-1} ( b_i^\dagger b_{i+1} + b_{i+1}^\dagger b_i)
\label{eq:HBS}
\end{equation}
The model contains a single Lindblad term which acts as a particle source at the center (site $i=0$) of the chain, at a rate parameterized by $\Gamma$:\footnote{Compared with Eq.~2 of Ref.~\cite{krapivsky_free_2020}, the factor 2 in the equation below comes from a different normalization used in their Lindblad equation.}
\begin{equation}
  L_0=\sqrt{2\Gamma}\, b_0^{\dagger}.
  \label{eq:LBS}
\end{equation}
Coding such a model can be done by defining the following \verb"evolver":
\begin{verbatim}
  hamiltonian(n) = sum(A(i)dag(A)(i+1)+dag(A)(i)A(i+1) for i in 1:n-1)
  dissipators(n, Gamma) = Dissipator(sqrt(2Gamma) * dag(A))(div(n, 2))
\end{verbatim}
This model is quasi-free and has been studied analytically by Krapivsky {\it et al.}~\cite{krapivsky_free_2020} in the case where the chain is empty at $t=0$.
In dimension one, the model displays a phase transition separating  a regime ($\Gamma<2$) where the total number of bosons $N(t)=\sum_i \langle b^\dag_i b_i \rangle$ grows quadratically and a regime   ($\Gamma>2$) where $N(t)$ grows exponentially. We illustrate here the use of the TMS library to study the small $\Gamma$ regime.
To perform the simulation of this model one has to specify the maximum boson occupation of the sites. To set the maximum occupation to 4 throughout the system and to start with an empty state in a mixed state representation, one can write:
\begin{verbatim}
  CreateState(
    type = Mixed(),
    system = System(n, Boson(5))
    state =  "0"
  )
\end{verbatim}

In Fig.~\ref{fig:boson} the simulation results are compared with
the exact solution of the model. The exact solution is obtained by solving a set of $N^2$ linear differential equations for the quantities $\langle b^\dagger_i b_j \rangle$. These equations are given in Eq.~10 of \cite{krapivsky_free_2020}.
Two quantities are displayed in  Fig.~\ref{fig:boson}: the density profile  $\langle b^\dag_i b_i \rangle$ (top panel) at time $t=1$ and $t=5$, and the mean number of bosons $N(t)$ (bottom panel). This simulation is carried out using the W$^{\rm II}$ algorithm at order 4 with a time step $\tau=0.1$ and a maximum bond dimension $\chi=200$, and it appears to accurately describe the dynamics up to time $t\simeq 5$.
Due to the large local Hilbert space dimension of such a bosonic system, it is however not straightforward to obtain accurate results at longer times. For this reason, checking the asymptotic results derived analytically in~\cite{krapivsky_free_2020} would require much longer simulations (the present calculation for the bosonic model, up to time $T=5$, with $\chi=200$ and $dim=5$ already required 100 hours on CPU time with 8 threads running in parallel).

\subsection{Free fermions with a localized source}
\label{ssec:fermion_source}

The model considered in this section is the fermionic analog of the previous model. The Hamiltonian describes spinless fermions hopping on the chain
\begin{equation}
  H =  \sum_{i=-N/2+1}^{N/2-1} ( c_i^{\dagger}  c_{i+1} + c_{i+1}^\dagger  c_i)
  \label{eq:HFS}
  \end{equation}
  and the particle injection in the center of the chain is due to the following jump operator
  \begin{equation}
    L_0=\sqrt{2\Gamma}\, c_0^\dagger.
    \label{eq:LFS}
  \end{equation}

\begin{figure}
  \centering
  \includegraphics[width=0.8\textwidth]{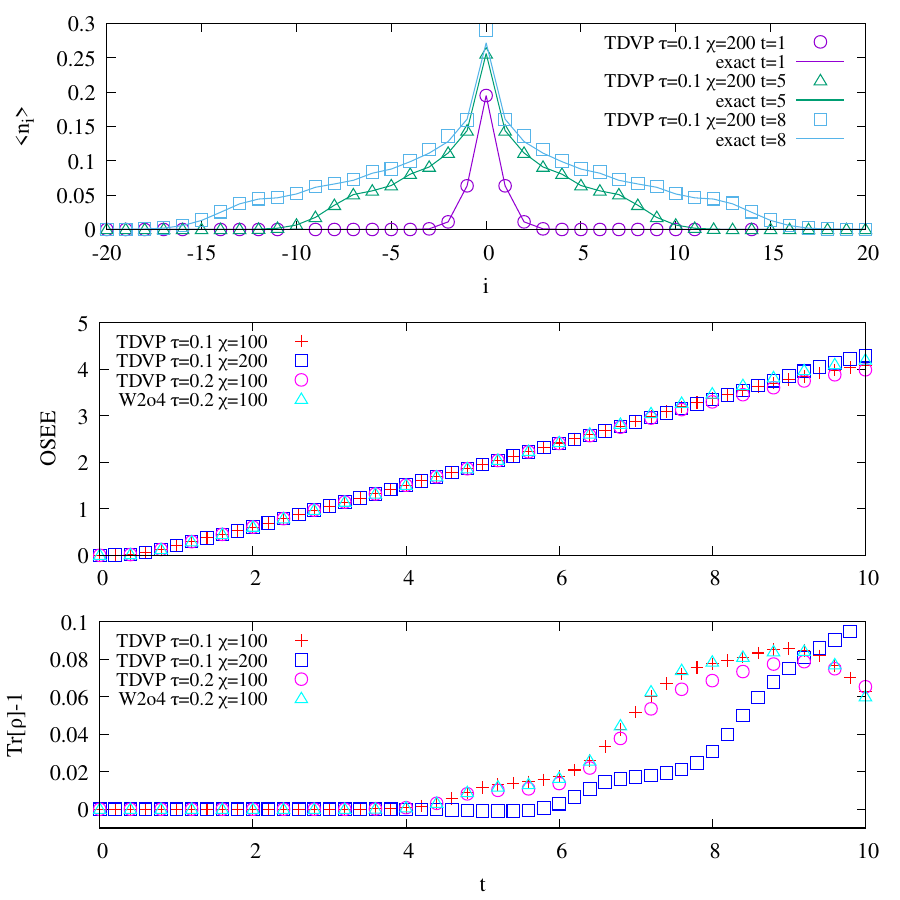}
  \caption{Free fermion model with a localized source (Eqs.~\ref{eq:HFS}-\ref{eq:LFS}). Top: density profile $\langle n_i(t)\rangle$ at three different times ($t=1$, 5 and 8). The full lines are exact results and the symbols have been obtained with TDVP with Trotter time step $\tau=0.1$, maximum bond dimension $\chi=200$. System size: $N=50$. At $t=1$ and $t=5$ the simulation reproduces almost perfectly the exact profiles. At time $t=8$ some discrepancy starts to be visible in the center of the profile and at the injection site $i=0$ in particular.
  Middle: linear growth of the OSEE taken in the middle bipartition. The different symbols correspond to simulations with different parameters or different algorithms (W$^{\rm II}$ at order 4 and TDVP). $\tau$ is the Trotter time step and $\chi$ the maximum bond dimension.
  Bottom: accumulated trace error as a function of time for different simulation parameters.}
  \label{fig:fermion_source}
\end{figure}

As for the model of Sec.~\ref{ssec:boson_source}, the model is quasi-free, and it has been studied analytically~\cite{krapivsky_free_2020}. The figure~\ref{fig:fermion_source} displays  the density profile at three different times,  the time evolution of the OSEE (for a bipartition in the center of the system), and the error on the trace of $\rho(t)$. The density profiles $\langle n_i(t)\rangle$ are compared with the exact solution.\footnote{This solution was obtained by solving numerically the set of $N^2$ differential equations describing the evolution of the two-point correlations ${\rm Tr}\left[\rho c^\dag_i c_j\right] $, see Eqs. 17 of Ref.~\cite{krapivsky_free_2019}. Up to signs these equations are very similar to those describing the dynamics of the 2-point correlations in the boson model of Sec.~\ref{ssec:boson_source}.}

The middle and the bottom panels of Fig.~\ref{fig:fermion_source} allow one to compare the precision of the TDVP and W$^{\rm II}$ and to see the influence of time step $\tau$ and maximum bond dimension $\chi$. In this example where the particle injection rate is $\Gamma=0.2$ all simulations are quantitatively accurate up to $t\simeq 5$. Beyond that time errors begin to be visible. Among the different simulations, the most accurate one is the one corresponding to $\chi=200$ with TDVP (blue squares in Fig.~\ref{fig:fermion_source}). Using an even larger bond dimension would allow one to describe the dynamics of the model at longer times.

Note that when an exact solution is not available, the error on the trace of $\rho$ (deviations from ${\rm Tr}\left[\rho\right]=1$) can be used to estimate at which time the simulations are no longer accurate enough.
For this particular problem the algorithms W$^{\rm II}$ at order 4 and TDVP turn out to offer a similar precision. It should be noted however that the execution time is much longer in the case of TDVP: the simulation up to time $T=8$ with W$^{\rm II}$ at order 4, $\tau=0.2$ and $\chi=200$ took 20 minutes on a CPU with 10 threads running in parallel, while the TDVP simulation with $\tau=0.1$ (giving a similar precision as W$^{\rm II}$ with $\tau=0.2$) took 340 minutes on the same processor.

\subsection{Decoherence of a complete-graph state}
\label{ssec:complete_graph}
The model presented in this section involves $N$ qubits that are initialized in a complete graph state and whose evolution is defined by a Lindblad equation only with dissipators (no Hamiltonian). For this model the time-dependence of numerous observables is known analytically, as discussed in Ref.~\cite{houdayer_solvable_2024}. 

\begin{figure}
  \centering
  \includegraphics[width=0.8\textwidth]{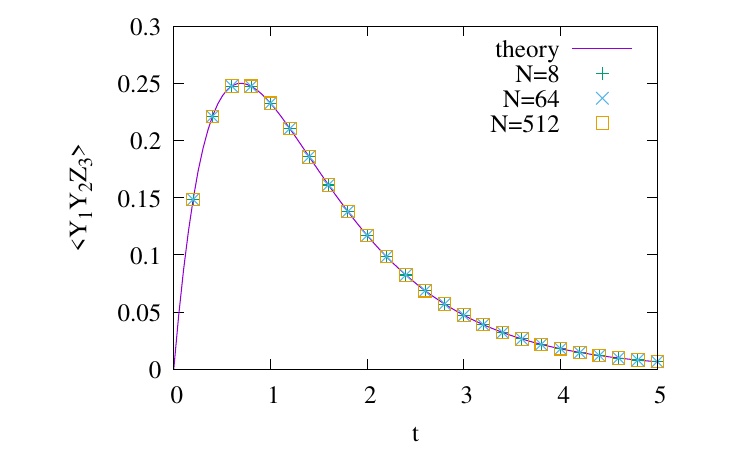}
  \caption{Time evolution of $\langle Y_1 Y_2 Z_3 \rangle$ in a model describing the decoherence of a complete-graph state for sizes 8, 64 and 512.
  The strength of the dissipation corresponds to $g_0=1.0$ in the notation of~\cite{houdayer_solvable_2024}.
  The solid line shows the exact result (Eq.~31 of~\cite{houdayer_solvable_2024}). Due to the permutation symmetry of this model the expectation value $\langle Y_i Y_j Z_k \rangle$ does not depend on $i,j$ and $k$ as long as they are different.}
  \label{fig:graph_state}
\end{figure}

A graph state~\cite{BR01,hein_entanglement_2006} is a pure and entangled state that is constructed by the application of controlled-Z (CZ) two-qubit gates to a product state:
\begin{equation}
  |g_E\rangle = \prod_{(i,j)\in E}{\rm CZ}(i,j) |++\cdots +\rangle,
\end{equation}
where $|+\rangle = \frac 1{\sqrt 2}(|0\rangle +|1\rangle)$ and the product runs over the edges of a graph $E$. Here we are dealing with the complete graph case where all possible edges are present in $E$. The construction of such an initial state (in pure representation) can be implemented as follows:
\begin{verbatim}
  state = create_graph_state(complete_graph(n))
\end{verbatim}
In the present model the dynamics is generated by the following dissipators
\begin{equation}
  L_i = \sigma^+_i \,,\,\, i=1\cdots N
\end{equation}
which are operators bringing the individual spins toward the state $|\uparrow\rangle$.
For this problem correlations turn out to be relatively low and the density matrix $\rho(t)$ can be represented by an MPS of bond dimension at most equal to 4~\cite{houdayer_solvable_2024}. TMS is thus able to manage very large systems (here we went up to $N=512$). Moreover, as the Lindbladian is only composed of single-site operators, the W$^{\rm II}$ approximation is exact and arbitrarily large time steps can be used without any Trotter error.\footnote{Note that since here the Lindbladian is a sum of single-site operators only, one could in principle construct the exact propagator $\exp(t\cal L)$ as a product of local operators.
The use of TMS with the W$^{\rm II}$ algorithm implements this automatically.}
Comparisons of numerical results produced by TMS with exact results from Ref.~\cite{houdayer_solvable_2024} are shown on Fig.~\ref{fig:graph_state} for a 3-site observable $\langle Y_1 Y_2 Z_3 \rangle$.

\subsection{Noisy quantum circuit}
\label{ssec:circuit}
We present here an example which illustrates how the library can be used to perform calculations on quantum circuits. This example is different from the previous ones since it is not associated to a continuous-time evolution. The circuit we consider for this example has a brick wall structure and is similar to the circuits encountered when
dealing with a Trotterized (discretized) Hamiltonian evolution.

\begin{figure}
  \centering
  \includegraphics[width=0.5\textwidth]{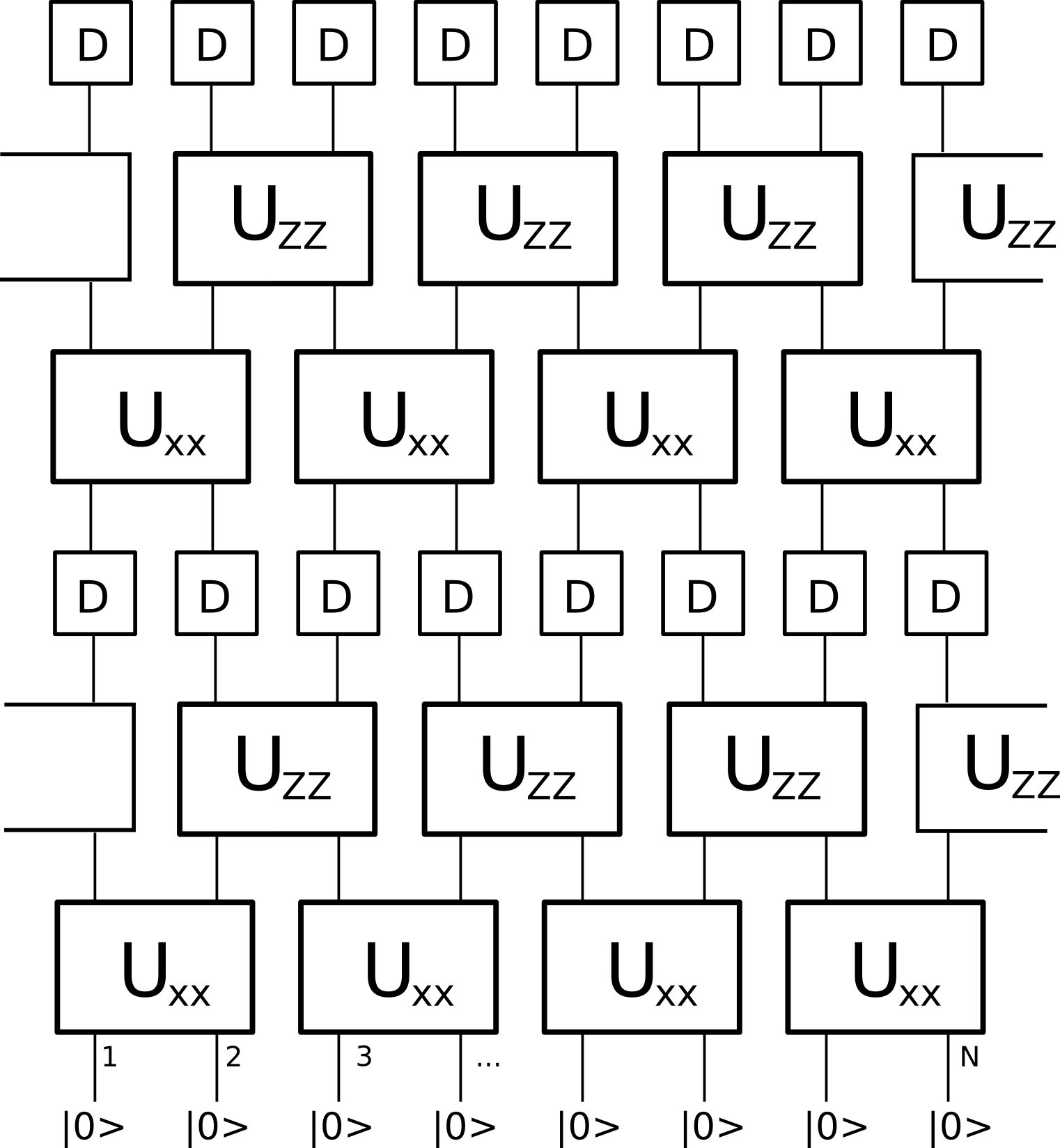}
  \caption{Brick wall quantum circuit made from layers of $U_{XX}^\phi(i,i+1)$ gates (Eq.~\ref{eq:lxx}), from layers of $U_{ZZ}^\phi(i,i+1)$ gates (Eq.~\ref{eq:lzz})
  and from layers of depolarization gates (Eq.~\ref{eq:epsilon}).}
  \label{fig:brickwallcircuit}
\end{figure}

Let us consider an even number of qubits $N$ and a quantum circuit that is built from successive layers of unitary two-qubit gates as well as layers representing
dissipative processes (or qubit errors).
The first circuit layer, $L_{XX}$, is unitary and is defined as a product of (commuting) 2-qubit gates:
\begin{equation}
  L_{XX} = U_{XX}^\phi(1,2) U_{XX}^\phi(3,4) \cdots U_{XX}^\phi(N-1,N)
  \label{eq:lxx}
\end{equation}
where $U_{XX}^\phi(i,j)$ acts on the qubits $i$ and $j$ and is defined by
\begin{equation}
  U_{XX}^\phi(i,j) =\exp(i\phi X_i X_j).
\end{equation}
The next layer, $L_{ZZ}$, is also defined by a product of (commuting) 2-qubit gates:
\begin{equation}
  L_{ZZ} = U_{ZZ}^\phi(N,1) U_{XX}^\phi(2,3) \cdots U_{XX}^\phi(N-2,N-1)
  \label{eq:lzz}
\end{equation}
where $U_{ZZ}^\phi(i,j)$ is defined by
\begin{equation}
  U_{ZZ}^\phi(i,j) =\exp(i\phi Z_i Z_j).
\end{equation}
The layers  $L_{XX}$ and $L_{ZZ}$ do not commute with each other and create entanglement.
After that we apply the gates that model qubit errors. Consider the depolarization channel:
\begin{equation}
  D_{i,p}: \rho \to \left(1-\frac{3}{4}p\right) \rho + \frac{1}{4}p X_i\rho X_i + \frac{1}{4}p Y_i\rho Y_i + \frac{1}{4}p Z_i\rho Z_i
  \label{eq:epsilon}
\end{equation}
where the parameter $0\leq p \leq 1$ represents an error probability.
The third layer of the circuit is the product of the  depolarization channels for all  qubits:
\begin{equation}
  L_\epsilon = \prod_i D_{i,p}
\end{equation}
Finally, the full circuit is a repetition $(L_\epsilon L_{ZZ} L_{XX}) (L_\epsilon L_{ZZ} L_{XX}) \cdots (L_\epsilon L_{ZZ} L_{XX})$
and the initial state is  all qubits in state $|0\rangle$. The circuit has a brick wall structure, as illustrated in Fig.~\ref{fig:brickwallcircuit}.

With TMS the needed operators are easily defined by
\begin{verbatim}
  # Depolarization channel
  DPL(p) = (1 - 0.75p) * Gate(Id) +
           0.25p * Gate(X) + 0.25p * Gate(Y) + 0.25p * Gate(Z)
  # Ising coupling gates
  Rxx(ϕ) = exp(-im * ϕ * X ⊗ X)
  Rzz(ϕ) = exp(-im * ϕ * Z ⊗ Z)
\end{verbatim}
And the gate sequence can be described by
\begin{verbatim}
[[Gates(
    name = "Applying exp(I*XX*ϕ) gates on qubits [1,2],[3,4],...",
    gates = prod(Rxx(ϕ)(2i-1,2i) for i in 1:div(n, 2)),
    limits = limits,
    ),
  Gates(
    name = "Applying exp(I*ZZ*ϕ) on qubits [N,1],[2,3],[4,5],...",
    gates = Rzz(ϕ)(1,n) *
            prod(Rzz(ϕ)(2i, 2i+1) for i in 1:(div(n,2))-1),
    limits = limits,
    ),
  Gates(
    name = "Depolarization channel on all qubits",
    final_measures = output(n),
    gates = prod(DPL(p)(i) for i in 1:n),
    limits = limits,
    )
] for _ in 1:steps]
\end{verbatim}

\begin{figure}
  \centering
  \includegraphics[width=0.8\textwidth]{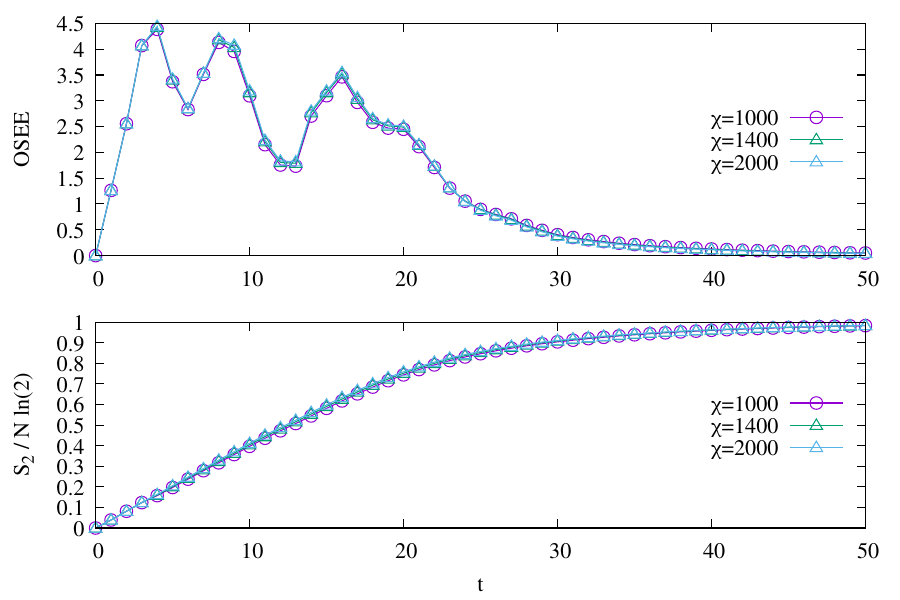}
  \caption{Circuit simulation. Top: OSEE as a function of the number $t$ of layers ($L_\epsilon L_{ZZ} L_{XX}$ counts as one complete layer). Bottom panel: second Rényi entropy $S_2$ normalized by its maximum value $N\ln 2$. System size: $N=20$. Error rate $p=0.02$ and gate angle $\phi=0.5$. Simulations with bond dimension $\chi=1000$, 1400 and 2000.}
  \label{fig:circuit}
\end{figure}

The top panel of Fig.~\ref{fig:circuit} represents the evolution of the OSEE for a partition in the center of the system as a function of the number of layers in the circuit. After an initial growth, due to the spread of correlations, the effect of the noise takes over when the number of layers becomes large. The state of the system then approaches an uncorrelated product state (with maximum Rényi-2 entropy). Due to the large amount of correlations generated by the initial layers of the circuit a relatively large bond dimension $\chi\sim 2000$ is required to get converged results.

\section{Conclusion}
\label{sec:conclusion}
We have presented TensorMixedStates, a Julia library for manipulating pure and mixed quantum states using
matrix product state representations. This library allows one in particular to apply unitary or non-unitary gates, as well as to solve continuous evolution equations such as the Schrödinger or the Lindblad equation. Based on ITensor, this library gives access to state-of-the-art algorithms such as TDVP or DMRG and MPS compression. Moreover, the particularly flexible and user-friendly interface allows simulations to be set up in a few lines of  code. We provided six examples to show the versatility and correctness of the software.
Five of the examples involved solving the Lindblad equation: two on fermions, one on bosons and two on spin 1/2. The last example demonstrated the use of non-unitary gates in a noisy quantum circuit calculation.

In the near future, we intend to work on several further developments. These include in particular: (i) the use of conserved quantum number (QNS) capabilities of ITensor to improve the efficiency and precision of certain simulations by taking into account conserved quantities of operators, and (ii) using automated MPO compression to improve the performance.

\section*{Acknowledgements}
We thank Haggai Landa for some previous collaborations on closely related topics. We also thank Pierre Cussenot for some discussions and feedback.

\paragraph{Funding information}
This work is supported by France 2030 under the French National Research Agency grant number ANR-22-QMET-0002 and by the PEPR integrated project EPiQ ANR-22-PETQ-0007.

\bibliography{tms}

@article{gorini_completely_1976,
	title = {Completely positive dynamical semigroups of {N}‐level systems},
	volume = {17},
	issn = {0022-2488},
	url = {https://doi.org/10.1063/1.522979},
	doi = {10.1063/1.522979},
	number = {5},
	journal = {J. Math. Phys.},
	author = {Gorini, Vittorio and Kossakowski, Andrzej and Sudarshan, E. C. G.},
	month = aug,
	year = {1976},
	pages = {821--825}
	}

@article{lindblad_generators_1976,
	title = {On the generators of quantum dynamical semigroups},
	volume = {48},
	issn = {1432-0916},
	url = {https://doi.org/10.1007/BF01608499},
	doi = {10.1007/BF01608499},
	language = {en},
	number = {2},
	journal = {Commun. Math. Phys.},
	author = {Lindblad, G.},
	month = jun,
	year = {1976},
	pages = {119--130}
}

@misc{ishiyama_exact_2025,
	title = {Exact density profile in a tight-binding chain with dephasing noise},
	url = {http://arxiv.org/abs/2501.07095},
	doi = {10.48550/arXiv.2501.07095},
	author = {Ishiyama, Taiki and Kazuya, Fujimoto and Sasamoto, Tomohiro},
	month = jan,
	year = {2025},
}

@article{yamanaka_exact_2023,
	title = {Exact solution for the {Lindbladian} dynamics for the open {XX} spin chain with boundary dissipation},
	volume = {14},
	issn = {2542-4653},
	url = {https://scipost.org/SciPostPhys.14.5.112},
	doi = {10.21468/SciPostPhys.14.5.112},
	language = {en},
	number = {5},
	journal = {SciPost Physics},
	author = {Yamanaka, Kohei and Sasamoto, Tomohiro},
	month = may,
	year = {2023},
	pages = {112},
}

@article{krapivsky_free_2020,
	title = {Free bosons with a localized source},
	volume = {2020},
	issn = {1742-5468},
	url = {https://dx.doi.org/10.1088/1742-5468/ab8118},
	doi = {10.1088/1742-5468/ab8118},
	language = {en},
	number = {6},
	journal = {J. Stat. Mech.},
	author = {Krapivsky, P L and Mallick, Kirone and Sels, Dries},
	month = jun,
	year = {2020},
	pages = {063101},
}

@article{barthel_solving_2022,
	title = {Solving quasi-free and quadratic {Lindblad} master equations for open fermionic and bosonic systems},
	volume = {2022},
	issn = {1742-5468},
	url = {https://dx.doi.org/10.1088/1742-5468/ac8e5c},
	doi = {10.1088/1742-5468/ac8e5c},
	language = {en},
	number = {11},
	journal = {J. Stat. Mech.},
	author = {Barthel, Thomas and Zhang, Yikang},
	month = nov,
	year = {2022},
	pages = {113101},
	}

@article{eisler_crossover_2011,
	title = {Crossover between ballistic and diffusive transport: the quantum exclusion process},
	volume = {2011},
	issn = {1742-5468},
	shorttitle = {Crossover between ballistic and diffusive transport},
	url = {https://dx.doi.org/10.1088/1742-5468/2011/06/P06007},
	doi = {10.1088/1742-5468/2011/06/P06007},
	language = {en},
	number = {06},
	journal = {J. Stat. Mech.},
	author = {Eisler, Viktor},
	month = jun,
	year = {2011},
	pages = {P06007},
}

@article{znidaric_exact_2010,
	title = {Exact solution for a diffusive nonequilibrium steady state of an open quantum chain},
	volume = {2010},
	issn = {1742-5468},
	url = {https://dx.doi.org/10.1088/1742-5468/2010/05/L05002},
	doi = {10.1088/1742-5468/2010/05/L05002},
	language = {en},
	number = {05},
	journal = {J. Stat. Mech.},
	author = {Žnidarič, Marko},
	month = may,
	year = {2010},
	pages = {L05002},
}

@article{zunkovic_closed_2014,
	title = {Closed hierarchy of correlations in {Markovian} open quantum systems},
	volume = {16},
	issn = {1367-2630},
	url = {https://dx.doi.org/10.1088/1367-2630/16/1/013042},
	doi = {10.1088/1367-2630/16/1/013042},
	language = {en},
	number = {1},
	journal = {New J. Phys.},
	author = {Žunkovič, Bojan},
	month = jan,
	year = {2014},
	pages = {013042},
}

@article{itensor,
	title={{The ITensor Software Library for Tensor Network Calculations}},
	author={Matthew Fishman and Steven R. White and E. Miles Stoudenmire},
	journal={SciPost Phys. Codebases},
	pages={4},
	year={2022},
	doi={10.21468/SciPostPhysCodeb.4},
	url={https://scipost.org/10.21468/SciPostPhysCodeb.4}
}

@article{prosen_third_2008,
	title = {Third quantization: a general method to solve master equations for quadratic open {Fermi} systems},
	volume = {10},
	issn = {1367-2630},
	shorttitle = {Third quantization},
	url = {https://dx.doi.org/10.1088/1367-2630/10/4/043026},
	doi = {10.1088/1367-2630/10/4/043026},
	language = {en},
	number = {4},
	journal = {New J. Phys.},
	author = {Prosen, Tomaž},
	month = apr,
	year = {2008},
	pages = {043026},
}

@article{krapivsky_free_2019,
	title = {Free fermions with a localized source},
	volume = {2019},
	issn = {1742-5468},
	url = {https://dx.doi.org/10.1088/1742-5468/ab4e8e},
	doi = {10.1088/1742-5468/ab4e8e},
	language = {en},
	number = {11},
	journal = {J. Stat. Mech.},
	author = {Krapivsky, P L and Mallick, Kirone and Sels, Dries},
	month = nov,
	year = {2019},
	pages = {113108},
}

@article{zaletel_time-evolving_2015,
	title = {Time-evolving a matrix product state with long-ranged interactions},
	volume = {91},
	url = {http://link.aps.org/doi/10.1103/PhysRevB.91.165112},
	doi = {10.1103/PhysRevB.91.165112},
	number = {16},
	journal = {Phys. Rev. B},
	author = {Zaletel, Michael P. and Mong, Roger S. K. and Karrasch, Christoph and Moore, Joel E. and Pollmann, Frank},
	month = apr,
	year = {2015},
	pages = {165112},
}

@article{bidzhiev_out--equilibrium_2017,
	title = {Out-of-equilibrium dynamics in a quantum impurity model: {Numerics} for particle transport and entanglement entropy},
	volume = {96},
	copyright = {All rights reserved},
	shorttitle = {Out-of-equilibrium dynamics in a quantum impurity model},
	url = {https://link.aps.org/doi/10.1103/PhysRevB.96.195117},
	doi = {10.1103/PhysRevB.96.195117},
	number = {19},
	journal = {Phys. Rev. B},
	author = {Bidzhiev, Kemal and Misguich, Grégoire},
	month = nov,
	year = {2017},
	pages = {195117},
}

@article{prosen_operator_2007,
	title = {Operator space entanglement entropy in a transverse {Ising} chain},
	volume = {76},
	url = {https://link.aps.org/doi/10.1103/PhysRevA.76.032316},
	doi = {10.1103/PhysRevA.76.032316},
	number = {3},
	journal = {Phys. Rev. A},
	author = {Prosen, Tomaž and Pižorn, Iztok},
	year = {2007},
	pages = {032316}
}

@article{znidaric_complexity_2008,
	title = {Complexity of thermal states in quantum spin chains},
	volume = {78},
	url = {https://link.aps.org/doi/10.1103/PhysRevA.78.022103},
	doi = {10.1103/PhysRevA.78.022103},
	number = {2},
	journal = {Phys. Rev. A},
	author = {Žnidarič, Marko and Prosen, Tomaž and Pižorn, Iztok},
	month = aug,
	year = {2008},
	pages = {022103},
}

@article{schollwock_density-matrix_2011,
	series = {January 2011 {Special} {Issue}},
	title = {The density-matrix renormalization group in the age of matrix product states},
	volume = {326},
	issn = {0003-4916},
	url = {http://www.sciencedirect.com/science/article/pii/S0003491610001752},
	doi = {10.1016/j.aop.2010.09.012},
	number = {1},
	urldate = {2014-05-06},
	journal = {Annals of Physics},
	author = {Schollwöck, Ulrich},
	month = jan,
	year = {2011},
	pages = {96--192},
}

@article{white_density_1992,
	title = {Density matrix formulation for quantum renormalization groups},
	volume = {69},
	url = {http://link.aps.org/doi/10.1103/PhysRevLett.69.2863},
	doi = {10.1103/PhysRevLett.69.2863},
	number = {19},
	journal = {Phys. Rev. Lett.},
	author = {White, Steven R.},
	month = nov,
	year = {1992},
	pages = {2863--2866},
}

@article{yang_time-dependent_2020,
	title = {Time-dependent variational principle with ancillary {Krylov} subspace},
	volume = {102},
	url = {https://link.aps.org/doi/10.1103/PhysRevB.102.094315},
	doi = {10.1103/PhysRevB.102.094315},
	number = {9},
	journal = {Phys. Rev. B},
	author = {Yang, Mingru and White, Steven R.},
	month = sep,
	year = {2020},
	pages = {094315}
}

@article{landa_nonlocal_2023,
	title={Nonlocal correlations in noisy multiqubit systems simulated using matrix product operators},
	author={Haggai Landa and Grégoire Misguich},
	journal={SciPost Phys. Core},
	volume={6},
	pages={037},
	year={2023},
	doi={10.21468/SciPostPhysCore.6.2.037},
	url={https://scipost.org/10.21468/SciPostPhysCore.6.2.037},
}

@misc{lindbladmpo,
  title={\href{https://github.com/qiskit-community/lindbladmpo}{github.com/qiskit-community/lindbladmpo}}
}

@misc{tensormixedstates,
  title={\href{https://github.com/jerhoud/TensorMixedStates.jl}{github.com/jerhoud/TensorMixedStates.jl}}
}

@misc{nielsen_quantum_2010,
	title = {Quantum {Computation} and {Quantum} {Information}: 10th {Anniversary} {Edition}},
	shorttitle = {Quantum {Computation} and {Quantum} {Information}},
	url = {https://www.cambridge.org/highereducation/books/quantum-computation-and-quantum-information/01E10196D0A682A6AEFFEA52D53BE9AE},
	language = {en},
	journal = {Higher Education from Cambridge University Press},
	author = {Nielsen, Michael A. and Chuang, Isaac L.},
	month = dec,
	year = {2010},
	doi = {10.1017/CBO9780511976667},
}

@article{weimer_simulation_2021,
	title = {Simulation methods for open quantum many-body systems},
	volume = {93},
	url = {https://link.aps.org/doi/10.1103/RevModPhys.93.015008},
	doi = {10.1103/RevModPhys.93.015008},
	number = {1},
	journal = {Rev. Mod. Phys.},
	author = {Weimer, Hendrik and Kshetrimayum, Augustine and Orús, Román},
	month = mar,
	year = {2021},
	pages = {015008},
}

@misc{fazio_many-body_2024,
	title = {Many-{Body} {Open} {Quantum} {Systems}},
	url = {http://arxiv.org/abs/2409.10300},
	doi = {10.48550/arXiv.2409.10300},
	author = {Fazio, Rosario and Keeling, Jonathan and Mazza, Leonardo and Schirò, Marco},
	month = oct,
	year = {2024},
}

@incollection{breuer_theory_2007,
	title = {The {Theory} of {Open} {Quantum} {Systems}},
	isbn = {978-0-19-921390-0},
	url = {https://doi.org/10.1093/acprof:oso/9780199213900.002.14006},
	booktitle = {The {Theory} of {Open} {Quantum} {Systems}},
	publisher = {Oxford University Press},
	author = {Breuer, Heinz-Peter and Petruccione, Francesco},
	month = jan,
	year = {2007},
	doi = {10.1093/acprof:oso/9780199213900.002.14006},
	pages = {0},
}

@article{zhou_what_2020,
	title = {What {Limits} the {Simulation} of {Quantum} {Computers}?},
	volume = {10},
	url = {https://link.aps.org/doi/10.1103/PhysRevX.10.041038},
	doi = {10.1103/PhysRevX.10.041038},
	number = {4},
	journal = {Phys. Rev. X},
	author = {Zhou, Yiqing and Stoudenmire, E. Miles and Waintal, Xavier},
	month = nov,
	year = {2020},
	pages = {041038},
}

@Article{tenpy2024,
    title={{Tensor network Python (TeNPy) version 1}},
    author={Johannes Hauschild and Jakob Unfried and Sajant Anand and Bartholomew Andrews and Marcus Bintz and Umberto Borla and Stefan Divic and Markus Drescher and Jan Geiger and Martin Hefel and Kévin Hémery and Wilhelm Kadow and Jack Kemp and Nico Kirchner and Vincent S. Liu and Gunnar Möller and Daniel Parker and Michael Rader and Anton Romen and Samuel Scalet and Leon Schoonderwoerd and Maximilian Schulz and Tomohiro Soejima and Philipp Thoma and Yantao Wu and Philip Zechmann and Ludwig Zweng and Roger S. K. Mong and Michael P. Zaletel and Frank Pollmann},
    journal={SciPost Phys. Codebases},
    pages={41},
    year={2024},
    doi={10.21468/SciPostPhysCodeb.41},
    url={https://scipost.org/10.21468/SciPostPhysCodeb.41},
}

@article{orus_tensor_2019,
	title = {Tensor networks for complex quantum systems},
	volume = {1},
	copyright = {2019 Springer Nature Limited},
	issn = {2522-5820},
	url = {https://www.nature.com/articles/s42254-019-0086-7},
	doi = {10.1038/s42254-019-0086-7},
	language = {en},
	number = {9},
	journal = {Nat Rev Phys},
	author = {Orús, Román},
	month = sep,
	year = {2019},
	pages = {538--550},
}

@article{begusic_fast_2024,
	title = {Fast and converged classical simulations of evidence for the utility of quantum computing before fault tolerance},
	volume = {10},
	url = {https://www.science.org/doi/10.1126/sciadv.adk4321},
	doi = {10.1126/sciadv.adk4321},
	number = {3},
	journal = {Science Advances},
	author = {Begušić, Tomislav and Gray, Johnnie and Chan, Garnet Kin-Lic},
	month = jan,
	year = {2024},
	pages = {eadk4321},
}

@article{stoudenmire_opening_2024,
	title = {Opening the {Black} {Box} inside {Grover}'s {Algorithm}},
	volume = {14},
	url = {https://link.aps.org/doi/10.1103/PhysRevX.14.041029},
	doi = {10.1103/PhysRevX.14.041029},
	number = {4},
	journal = {Phys. Rev. X},
	author = {Stoudenmire, E. M. and Waintal, Xavier},
	month = nov,
	year = {2024},
	pages = {041029},
}

@article{orus_practical_2014,
	title = {A practical introduction to tensor networks: {Matrix} product states and projected entangled pair states},
	volume = {349},
	issn = {0003-4916},
	shorttitle = {A practical introduction to tensor networks},
	url = {http://www.sciencedirect.com/science/article/pii/S0003491614001596},
	doi = {10.1016/j.aop.2014.06.013},
	journal = {Annals of Physics},
	author = {Orús, Román},
	month = oct,
	year = {2014},
	pages = {117--158},
}

@misc{king_computational_2024,
	title = {Computational supremacy in quantum simulation},
	url = {http://arxiv.org/abs/2403.00910},
	doi = {10.48550/arXiv.2403.00910},
	urldate = {2024-06-03},
	author = {King, Andrew D. and Nocera, Alberto and Rams, Marek M. and Dziarmaga, Jacek and Wiersema, Roeland and Bernoudy, William and Raymond, Jack and Kaushal, Nitin and Heinsdorf, Niclas and Harris, Richard and Boothby, Kelly and Altomare, Fabio and Berkley, Andrew J. and Boschnak, Martin and Chern, Kevin and Christiani, Holly and Cibere, Samantha and Connor, Jake and Dehn, Martin H. and Deshpande, Rahul and Ejtemaee, Sara and Farré, Pau and Hamer, Kelsey and Hoskinson, Emile and Huang, Shuiyuan and Johnson, Mark W. and Kortas, Samuel and Ladizinsky, Eric and Lai, Tony and Lanting, Trevor and Li, Ryan and MacDonald, Allison J. R. and Marsden, Gaelen and McGeoch, Catherine C. and Molavi, Reza and Neufeld, Richard and Norouzpour, Mana and Oh, Travis and Pasvolsky, Joel and Poitras, Patrick and Poulin-Lamarre, Gabriel and Prescott, Thomas and Reis, Mauricio and Rich, Chris and Samani, Mohammad and Sheldan, Benjamin and Smirnov, Anatoly and Sterpka, Edward and Clavera, Berta Trullas and Tsai, Nicholas and Volkmann, Mark and Whiticar, Alexander and Whittaker, Jed D. and Wilkinson, Warren and Yao, Jason and Yi, T. J. and Sandvik, Anders W. and Alvarez, Gonzalo and Melko, Roger G. and Carrasquilla, Juan and Franz, Marcel and Amin, Mohammad H.},
	month = mar,
	year = {2024},
}

@article{ayral_density-matrix_2023,
	title = {Density-{Matrix} {Renormalization} {Group} {Algorithm} for {Simulating} {Quantum} {Circuits} with a {Finite} {Fidelity}},
	volume = {4},
	url = {https://link.aps.org/doi/10.1103/PRXQuantum.4.020304},
	doi = {10.1103/PRXQuantum.4.020304},
	number = {2},
	journal = {PRX Quantum},
	author = {Ayral, Thomas and Louvet, Thibaud and Zhou, Yiqing and Lambert, Cyprien and Stoudenmire, E. Miles and Waintal, Xavier},
	month = apr,
	year = {2023},
	pages = {020304},
}

@article{verstraete_matrix_2004,
	title = {Matrix {Product} {Density} {Operators}: {Simulation} of {Finite}-{Temperature} and {Dissipative} {Systems}},
	volume = {93},
	shorttitle = {Matrix {Product} {Density} {Operators}},
	url = {https://link.aps.org/doi/10.1103/PhysRevLett.93.207204},
	doi = {10.1103/PhysRevLett.93.207204},
	number = {20},
	journal = {Phys. Rev. Lett.},
	author = {Verstraete, F. and García-Ripoll, J. J. and Cirac, J. I.},
	month = nov,
	year = {2004},
	pages = {207204},
}

@article{cirac_matrix_2021,
	title = {Matrix product states and projected entangled pair states: {Concepts}, symmetries, theorems},
	volume = {93},
	shorttitle = {Matrix product states and projected entangled pair states},
	url = {https://link.aps.org/doi/10.1103/RevModPhys.93.045003},
	doi = {10.1103/RevModPhys.93.045003},
	number = {4},
	journal = {Rev. Mod. Phys.},
	author = {Cirac, J. Ignacio and Pérez-García, David and Schuch, Norbert and Verstraete, Frank},
	month = dec,
	year = {2021},
	pages = {045003},
}

@article{houdayer_solvable_2024,
	title = {A solvable model for graph state decoherence dynamics},
	volume = {7},
	issn = {2666-9366},
	url = {https://scipost.org/SciPostPhysCore.7.1.009},
	doi = {10.21468/SciPostPhysCore.7.1.009},
	language = {en},
	number = {1},
	urldate = {2025-05-14},
	journal = {SciPost Physics Core},
	author = {Houdayer, Jérôme and Landa, Haggai and Misguich, Grégoire},
	month = feb,
	year = {2024},
	pages = {009},
}

@article{lacroix_mpsdynamicsjl_2024,
	title = {{MPSDynamics}.jl: {Tensor} network simulations for finite-temperature (non-{Markovian}) open quantum system dynamics},
	volume = {161},
	issn = {0021-9606},
	shorttitle = {{MPSDynamics}.jl},
	url = {https://doi.org/10.1063/5.0223107},
	doi = {10.1063/5.0223107},
	number = {8},
	journal = {The Journal of Chemical Physics},
	author = {Lacroix, Thibaut and Le Dé, Brieuc and Riva, Angela and Dunnett, Angus J. and Chin, Alex W.},
	month = aug,
	year = {2024},
	pages = {084116},
}

@article{gray2018quimb,
    title={quimb: a python library for quantum information and many-body calculations},
    author={Gray, Johnnie},
    journal={Journal of Open Source Software},
    year = {2018},
    volume={3}, number={29}, pages={819},
    doi={10.21105/joss.00819},
}

@article{pastaq,
    title={\mbox{PastaQ}: A Package for Simulation, Tomography and Analysis of Quantum Computers},
    author={Giacomo Torlai and Matthew Fishman},
    year={2020},
    url={https://github.com/GTorlai/PastaQ.jl/},
	journal={\href{https://github.com/GTorlai/PastaQ.jl}{github.com/GTorlai/PastaQ.jl}}
}

@article{BR01,
  title = {Persistent Entanglement in Arrays of Interacting Particles},
  author = {Briegel, Hans J. and Raussendorf, Robert},
  journal = {Phys. Rev. Lett.},
  volume = {86},
  issue = {5},
  pages = {910--913},
  numpages = {0},
  year = {2001},
  month = {Jan},
  doi = {10.1103/PhysRevLett.86.910},
  url = {https://link.aps.org/doi/10.1103/PhysRevLett.86.910}
}

@incollection{hein_entanglement_2006,
	title = {Entanglement in graph states and its applications},
	url = {https://ebooks.iospress.nl/doi/10.3254/978-1-61499-018-5-115},
	urldate = {2023-05-19},
	booktitle = {Quantum {Computers}, {Algorithms} and {Chaos}},
	publisher = {IOS Press},
	author = {Hein, M. and Dür, W. and Eisert, J. and Raussendorf, R. and Van den Nest, M. and Briegel, H.-J.},
	year = {2006},
	doi = {10.3254/978-1-61499-018-5-115},
	pages = {115--218},
}

@article{blattQuantumSimulationsTrapped2012,
  title = {Quantum Simulations with Trapped Ions},
  author = {Blatt, R. and Roos, C. F.},
  year = {2012},
  month = apr,
  journal = {Nature Physics},
  volume = {8},
  number = {4},
  pages = {277--284},
  issn = {1745-2481},
  doi = {10.1038/nphys2252},
  langid = {english}
}

@article{bruzewiczTrappedIonQuantumComputing2019,
  title = {Trapped-{{Ion Quantum Computing}}: {{Progress}} and {{Challenges}}},
  shorttitle = {Trapped-{{Ion Quantum Computing}}},
  author = {Bruzewicz, Colin D. and Chiaverini, John and McConnell, Robert and Sage, Jeremy M.},
  year = {2019},
  month = jun,
  journal = {Applied Physics Reviews},
  volume = {6},
  number = {2},
  eprint = {1904.04178},
  primaryclass = {quant-ph},
  pages = {021314},
  issn = {1931-9401},
  doi = {10.1063/1.5088164},
  urldate = {2025-09-29},
}

@article{blochQuantumSimulationsUltracold2012,
  title = {Quantum Simulations with Ultracold Quantum Gases},
  author = {Bloch, Immanuel and Dalibard, Jean and Nascimb{\`e}ne, Sylvain},
  year = {2012},
  month = apr,
  journal = {Nature Physics},
  volume = {8},
  number = {4},
  pages = {267--276},
  issn = {1745-2481},
  doi = {10.1038/nphys2259},
  urldate = {2025-08-29},
  copyright = {2012 Springer Nature Limited},
  langid = {english}
}

@article{devoret_superconducting_2013,
	title = {Superconducting circuits for quantum information: an outlook},
	volume = {16},
	url = {https://www.science.org/doi/10.1126/science.1231930},
	doi = {10.1126/science.1231930},
	number = {5364},
	journal = {Science},
	author = {Devoret, Michel H. and Schoelkopf, Robert J.},
	month = apr,
	year = {2013},
	pages = {1169--1174},
}

@article{saffmanQuantumInformationRydberg2010,
  title = {Quantum Information with {{Rydberg}} Atoms},
  author = {Saffman, M. and Walker, T. G. and M{\o}lmer, K.},
  year = {2010},
  month = aug,
  journal = {Rev.Mod.Phys.},
  volume = {82},
  number = {3},
  pages = {2313--2363},
  publisher = {American Physical Society},
  doi = {10.1103/RevModPhys.82.2313}
}

@article{bernienProbingManybodyDynamics2017,
  title = {Probing Many-Body Dynamics on a 51-Atom Quantum Simulator},
  author = {Bernien, Hannes and Schwartz, Sylvain and Keesling, Alexander and Levine, Harry and Omran, Ahmed and Pichler, Hannes and Choi, Soonwon and Zibrov, Alexander S. and Endres, Manuel and Greiner, Markus and Vuleti{\'c}, Vladan and Lukin, Mikhail D.},
  year = {2017},
  month = nov,
  journal = {Nature},
  volume = {551},
  number = {7682},
  pages = {579--584},
  publisher = {Nature Publishing Group},
  issn = {1476-4687},
  doi = {10.1038/nature24622},
  langid = {english}
}

@article{browaeysManybodyPhysicsIndividually2020,
  title = {Many-Body Physics with Individually Controlled {{Rydberg}} Atoms},
  author = {Browaeys, Antoine and Lahaye, Thierry},
  year = {2020},
  month = feb,
  journal = {Nature Physics},
  volume = {16},
  number = {2},
  pages = {132--142},
  publisher = {Nature Publishing Group},
  issn = {1745-2481},
  doi = {10.1038/s41567-019-0733-z},
  copyright = {2020 Springer Nature Limited},
  langid = {english}
}

@article{preskillQuantumComputingNISQ2018,
  title = {Quantum {{Computing}} in the {{NISQ}} Era and Beyond},
  author = {Preskill, John},
  year = {2018},
  month = aug,
  journal = {Quantum},
  volume = {2},
  pages = {79},
  publisher = {Verein zur F{\"o}rderung des Open Access Publizierens in den Quantenwissenschaften},
  doi = {10.22331/q-2018-08-06-79},
  langid = {british},
}

@article{monroeProgrammableQuantumSimulations2021,
  title = {Programmable Quantum Simulations of Spin Systems with Trapped Ions},
  author = {Monroe, C. and Campbell, W. C. and Duan, L.-M. and Gong, Z.-X. and Gorshkov, A. V. and Hess, P. W. and Islam, R. and Kim, K. and Linke, N. M. and Pagano, G. and Richerme, P. and Senko, C. and Yao, N. Y.},
  year = {2021},
  month = apr,
  journal = {Reviews of Modern Physics},
  volume = {93},
  number = {2},
  pages = {025001},
  publisher = {American Physical Society},
  doi = {10.1103/RevModPhys.93.025001},
 }

@article{MORVAN_PhaseTransitionsRandom_2024,
  title = {Phase Transitions in Random Circuit Sampling},
  author = {Morvan, A. and Villalonga, B. and Mi, X. and Mandr{\`a}, S. and Bengtsson, A. and Klimov, P. V. and Chen, Z. and Hong, S. and Erickson, C. and Drozdov, I. K. and Chau, J. and Laun, G. and Movassagh, R. and Asfaw, A. and Brand{\~a}o, L. T. a. N. and Peralta, R. and Abanin, D. and Acharya, R. and Allen, R. and Andersen, T. I. and Anderson, K. and Ansmann, M. and Arute, F. and Arya, K. and Atalaya, J. and Bardin, J. C. and Bilmes, A. and Bortoli, G. and Bourassa, A. and Bovaird, J. and Brill, L. and Broughton, M. and Buckley, B. B. and Buell, D. A. and Burger, T. and Burkett, B. and Bushnell, N. and Campero, J. and Chang, H.-S. and Chiaro, B. and Chik, D. and Chou, C. and Cogan, J. and Collins, R. and Conner, P. and Courtney, W. and Crook, A. L. and Curtin, B. and Debroy, D. M. and Barba, A. Del Toro and Demura, S. and Paolo, A. Di and Dunsworth, A. and Faoro, L. and Farhi, E. and Fatemi, R. and Ferreira, V. S. and Burgos, L. Flores and Forati, E. and Fowler, A. G. and Foxen, B. and Garcia, G. and Genois, {\'E} and Giang, W. and Gidney, C. and Gilboa, D. and Giustina, M. and Gosula, R. and Dau, A. Grajales and Gross, J. A. and Habegger, S. and Hamilton, M. C. and Hansen, M. and Harrigan, M. P. and Harrington, S. D. and Heu, P. and Hoffmann, M. R. and Huang, T. and Huff, A. and Huggins, W. J. and Ioffe, L. B. and Isakov, S. V. and Iveland, J. and Jeffrey, E. and Jiang, Z. and Jones, C. and Juhas, P. and Kafri, D. and Khattar, T. and Khezri, M. and Kieferov{\'a}, M. and Kim, S. and Kitaev, A. and Klots, A. R. and Korotkov, A. N. and Kostritsa, F. and Kreikebaum, J. M. and Landhuis, D. and Laptev, P. and Lau, K.-M. and Laws, L. and Lee, J. and Lee, K. W. and Lensky, Y. D. and Lester, B. J. and Lill, A. T. and Liu, W. and Livingston, W. P. and Locharla, A. and Malone, F. D. and Martin, O. and Martin, S. and McClean, J. R. and McEwen, M. and Miao, K. C. and Mieszala, A. and Montazeri, S. and Mruczkiewicz, W. and Naaman, O. and Neeley, M. and Neill, C. and Nersisyan, A. and Newman, M. and Ng, J. H. and Nguyen, A. and Nguyen, M. and Niu, M. Yuezhen and O'Brien, T. E. and Omonije, S. and Opremcak, A. and Petukhov, A. and Potter, R. and Pryadko, L. P. and Quintana, C. and Rhodes, D. M. and Rocque, C. and Rosenberg, E. and Rubin, N. C. and Saei, N. and Sank, D. and Sankaragomathi, K. and Satzinger, K. J. and Schurkus, H. F. and Schuster, C. and Shearn, M. J. and Shorter, A. and Shutty, N. and Shvarts, V. and Sivak, V. and Skruzny, J. and Smith, W. C. and Somma, R. D. and Sterling, G. and Strain, D. and Szalay, M. and Thor, D. and Torres, A. and Vidal, G. and Heidweiller, C. Vollgraff and White, T. and Woo, B. W. K. and Xing, C. and Yao, Z. J. and Yeh, P. and Yoo, J. and Young, G. and Zalcman, A. and Zhang, Y. and Zhu, N. and Zobrist, N. and Rieffel, E. G. and Biswas, R. and Babbush, R. and Bacon, D. and Hilton, J. and Lucero, E. and Neven, H. and Megrant, A. and Kelly, J. and Roushan, P. and Aleiner, I. and Smelyanskiy, V. and Kechedzhi, K. and Chen, Y. and Boixo, S.},
  year = {2024},
  month = oct,
  journal = {Nature},
  volume = {634},
  number = {8033},
  pages = {328--333},
  publisher = {Nature Publishing Group},
  issn = {1476-4687},
  doi = {10.1038/s41586-024-07998-6},
  langid = {english},
}

@article{ARUTE_QuantumSupremacyUsing_2019,
  title = {Quantum Supremacy Using a Programmable Superconducting Processor},
  author = {Arute, Frank and Arya, Kunal and Babbush, Ryan and Bacon, Dave and Bardin, Joseph C. and Barends, Rami and Biswas, Rupak and Boixo, Sergio and Brandao, Fernando G. S. L. and Buell, David A. and Burkett, Brian and Chen, Yu and Chen, Zijun and Chiaro, Ben and Collins, Roberto and Courtney, William and Dunsworth, Andrew and Farhi, Edward and Foxen, Brooks and Fowler, Austin and Gidney, Craig and Giustina, Marissa and Graff, Rob and Guerin, Keith and Habegger, Steve and Harrigan, Matthew P. and Hartmann, Michael J. and Ho, Alan and Hoffmann, Markus and Huang, Trent and Humble, Travis S. and Isakov, Sergei V. and Jeffrey, Evan and Jiang, Zhang and Kafri, Dvir and Kechedzhi, Kostyantyn and Kelly, Julian and Klimov, Paul V. and Knysh, Sergey and Korotkov, Alexander and Kostritsa, Fedor and Landhuis, David and Lindmark, Mike and Lucero, Erik and Lyakh, Dmitry and Mandr{\`a}, Salvatore and McClean, Jarrod R. and McEwen, Matthew and Megrant, Anthony and Mi, Xiao and Michielsen, Kristel and Mohseni, Masoud and Mutus, Josh and Naaman, Ofer and Neeley, Matthew and Neill, Charles and Niu, Murphy Yuezhen and Ostby, Eric and Petukhov, Andre and Platt, John C. and Quintana, Chris and Rieffel, Eleanor G. and Roushan, Pedram and Rubin, Nicholas C. and Sank, Daniel and Satzinger, Kevin J. and Smelyanskiy, Vadim and Sung, Kevin J. and Trevithick, Matthew D. and Vainsencher, Amit and Villalonga, Benjamin and White, Theodore and Yao, Z. Jamie and Yeh, Ping and Zalcman, Adam and Neven, Hartmut and Martinis, John M.},
  year = {2019},
  month = oct,
  journal = {Nature},
  volume = {574},
  number = {7779},
  pages = {505--510},
  publisher = {Nature Publishing Group},
  issn = {1476-4687},
  doi = {10.1038/s41586-019-1666-5},
  urldate = {2025-09-30},
  copyright = {2019 The Author(s), under exclusive licence to Springer Nature Limited},
  langid = {english},
}

\end{document}